# A Self-Calibrating Framework for Analog Circuit Sizing Using LLM-Derived Analytical Equations

Antonio J. Bujana[1,2], *Graduate Student Member, IEEE,* and Aydin I. Karsilayan[1], *Member, IEEE*

***Abstract*— We present a design automation framework for analog circuit sizing that produces calibrated, topology-specific analytical equations from raw circuit netlists. A large language model (LLM) derives a complete Python sizing function in which each device dimension is traceable to a specific design rationale — a form of interpretable output absent from existing optimization-based and LLM-based sizing methods. A deterministic calibration loop extracts process-dependent parameters from a single DC operating point simulation, while a prediction-error feedback mechanism compensates for analytical inaccuracies. We validate the framework on circuits ranging from 6 to 30 transistors — spanning single-stage, current-mirror (simple and cascoded), folded-cascode, gain-boosted folded-cascode, two-stage Miller-compensated, nested-Miller-compensated, and complementary class-AB output topologies — across six process nodes from 32 nm to 180 nm. On matched-specification benchmarks, including the class-AB opamp case, the framework converges within a few simulations. Despite large initial prediction errors, convergence depends on the measurement-feedback architecture, not prediction accuracy. The one-shot calibration automatically captures process-dependent variations, enabling cross-node portability without modification, retraining, or per-process characterization.**

## I. INTRODUCTION

ANALOG circuit sizing remains a labor-intensive task requiring expert knowledge to navigate complex tradeoffs among competing performance metrics — gain, bandwidth (gain-bandwidth product, GBW), phase margin (PM), slew rate (SR), and power consumption. Despite decades of research into automated sizing — spanning optimization-based, knowledge-based, and large language model (LLM)-based methods (reviewed in Section II) — to the authors' knowledge, no existing approach produces executable, topology-specific design equations directly from raw circuit netlists, and no prior work provides quantitative analysis of how prediction accuracy relates to convergence.

This paper introduces a fundamentally different role for the LLM: deriving the analytical design equations themselves. The LLM receives a raw circuit netlist and produces a complete Python sizing function (`compute_sizing()`) encoding topology-specific equations — equations consistent with standard analog design theory, but generated automatically from the netlist. A deterministic calibration loop then anchors these equations in process-accurate simulation data by extracting process-dependent parameters from a single DC operating point (OP) simulation. The key insight is that LLM-generated equations need only be structurally correct — even if their numerical predictions are inaccurate. We define structural correctness as the property that the equations (i) correctly identify which devices govern each specification, (ii) correctly capture the direction of each device's influence, and (iii) correctly represent how specifications trade off against each other through shared devices. Structural correctness is essential: when a specification falls short, structurally correct equations direct the LLM to adjust the right devices in the right direction — even if the quantitative adjustment is imprecise — so the prediction-error feedback drives the design toward convergence rather than oscillation. The calibration loop corrects numerical errors, and the prediction-error feedback compensates for systematic analytical inaccuracies that persist after calibration. We validate the framework against existing methods on matched circuits and specifications, and analyze topology-dependent convergence behavior across process nodes.

The main contributions of this work are as follows:

1) An end-to-end, self-calibrating sizing workflow validated on circuits ranging from 6 to 30 transistors across eight topology families, including single-stage, two-stage, and three-stage compensation architectures. Unlike existing LLM-based sizing approaches that produce only final device dimensions, this framework preserves calibrated, human-readable design equations — commented Python code tracing each device dimension to a specific design rationale. The only per-circuit inputs are the netlist, target specifications, and supply voltages; all scripts and prompt templates are shared across topologies and nodes with no modification.
2) A calibration feedback architecture that decouples convergence from the LLM's numerical prediction accuracy. The LLM-derived equations must be structurally correct — correctly mapping specifications to governing devices and influence directions — but need not be numerically accurate. Quantitative prediction-error analysis confirms this decoupling (Section V-B).
3) Cross-node portability via self-calibration. The one-shot DC OP extraction automatically captures process-

[1]Department of Electrical and Computer Engineering, Texas A&M University, College Station, TX, USA. [2]Department of Engineering, Benedictine College, Atchison, KS, USA. Correspondence to: A. J. Bujana tonybujana@tamu.edu

dependent parameters ($\mu C_{ox}$, $\alpha g_m$, $\lambda$, $V_{th}$) that vary significantly across process nodes, enabling the same framework to self-calibrate across all six nodes (32 nm to 180 nm) without modification, per-process characterization data, or retraining.

## II. Prior Work

Existing automated sizing approaches fall into three broad categories: optimization-based methods that treat the circuit as a black box, knowledge-based methods that rely on pre-characterized device data, and LLM-based methods that leverage language models in various roles. Each is reviewed below.

### A. Optimization-Based Methods

Black-box optimization methods formulate circuit sizing as a numerical optimization problem. Bayesian optimization (BO) [1] uses Gaussian process surrogates to efficiently explore the design space. Traditional implementations require hundreds to thousands of simulations to converge, though recent work on initialization strategies [2] has substantially reduced this for OTA-class circuits, achieving design closure in as few as 21 total simulations. Geometric programming [3] achieves faster convergence through convex formulations but requires the designer to manually construct posynomial models for each topology. Reinforcement learning methods [4] train agents through simulation rewards but require per-topology training episodes and produce no analytical insight. These methods achieve good performance but treat the circuit as a black box — the final sizing comes with no interpretable rationale explaining why specific dimensions were chosen.

### B. Knowledge-Based Methods

The $g_m/I_D$ methodology [5] represents the most widely adopted knowledge-based approach. It uses pre-characterized lookup tables relating transconductance efficiency to inversion level, enabling systematic operating-point selection. However, it requires per-process characterization sweeps (typically $W$ and $L$ sweeps across the full operating range), and a human expert must interpret efficiency curves, select operating points, and manually derive the design equations linking specifications to device dimensions. These two requirements — manual equation derivation and per-process characterization — represent the primary bottlenecks of the $g_m/I_D$ approach.

### C. LLM-Based Sizing Methods

The landscape of LLM-based analog sizing methods has expanded rapidly. These approaches can be categorized by the role assigned to the LLM.

*1) LLM as optimizer:* EEsizer [6] uses an LLM directly to propose sizing adjustments based on simulation feedback. The LLM analyzes performance gaps and suggests parameter changes iteratively. Testing across eight LLMs on basic circuits, the authors scaled to a 20-transistor opamp and demonstrated cross-node applicability across 180 nm, 130 nm, and 90 nm, achieving target specifications across multiple test groups.

*2) LLM as search-space guide:* LEDRO [7] uses an LLM to iteratively reduce the design search space for a TuRBO optimizer. By having the LLM propose refined parameter ranges based on simulation feedback, LEDRO reduces the exploration complexity for the downstream optimizer. Tested on 22 op-amp topologies across four FinFET nodes, it demonstrates broad generalizability. However, convergence ultimately depends on the TuRBO optimizer, and the LLM's role is limited to search-space pruning rather than analytical reasoning about circuit behavior.

*3) LLM as knowledge transfer agent:* LLM-USO [8] introduces structured knowledge representations encoding relationships between performance metrics, circuit sub-structures, and sizing parameters, enabling transfer learning across circuits with similar sub-structures via a knowledge summary mechanism at 45 simulations per circuit. ADO-KT [9] provides a complementary form of structure — LLM-generated knowledge graphs of reusable relationships among specifications, subcircuits, and sizing parameters, enabling topology-agnostic transfer of prior design knowledge. ADO-LLM [10] combines in-context learning with BO, using the LLM to propose promising design points within a BO framework.

*4) LLM as multi-agent system:* AnaFlow [11] employs specialized LLM agents that collaborate in a multi-agent workflow, providing structured reasoning and explainability. Atelier [12] extends this with a graph-of-thoughts architecture integrating a curated knowledge base, achieving superior performance over black-box methods on specific topologies. The framework uses 100–400 simulation evaluations per sizing. AutoSizer [13] introduces a reflective meta-optimization framework where LLM agents adaptively select and configure optimization algorithms within a two-loop architecture, tested on 24 circuits. TopoSizing [14] performs circuit understanding from raw netlists using graph algorithms to organize circuits into hierarchical device–module–stage representations, then uses LLM agents in a hypothesis-verification loop to produce annotations that guide BO.

*5) Lightweight and cross-node methods:* EasySize [15] fine-tunes a lightweight Qwen3-8B model for cross-node applicability, achieving strong performance at 180 nm, 45 nm, and 22 nm nodes despite training only on 350 nm data. LLMACD [16] operates in transistor behavioral-parameter space, using a behavioral model to map LLM-generated parameters to device sizes with manually embedded design knowledge. AmpAgent [17] uses literature retrieval to port amplifier designs across processes and performance targets.

## III. Proposed Methodology

Despite the breadth of existing approaches (Section II), LLM-based sizing methods share a common architectural pattern: the LLM serves as either an optimizer — suggesting parameter adjustments based on simulation feedback — or a search-space guide that reduces the region for a downstream optimizer. Table I positions the proposed framework against these methods; the key differentiators are detailed in the contributions (Section I).

Existing methods that claim explainability (AnaFlow, Atelier) provide reasoning traces explaining why the LLM made specific parameter adjustments — valuable for understanding the optimization trajectory, but not equivalent to calibrated design equations — and TopoSizing [14] produces structured annotations but uses them to guide BO rather than deriving executable sizing equations. The proposed framework outputs a complete sizing function where each line of code maps a specification to device dimensions through explicit analytical relationships. Beyond explainability, this provides traceability for design review — before committing to layout, a designer can inspect which specification constrained each device dimension — and diagnostic transparency: when convergence is slow or a specification is marginal, the calibrated equations identify whether the root cause is a parametric error (e.g., inaccurate $\mu C_{ox}$) or a fundamental topological limitation. As discussed in Section VI-A, the equations also provide the LLM with a causal model of the circuit that anchors its sizing adjustments across rounds.

### A. Framework Overview

The proposed framework is a simulation-in-the-loop sizing method: the analytical layer proposes and explains each round's design, and every accepted result is established by measurement. It operates in iterative rounds, as shown in Fig. 1, each comprising four steps: prompt assembly, LLM equation generation or update, sizing execution, and simulation with calibration.

*1) Round 0:* A prompt-assembly function, `build_prompt()`, combines the circuit netlist, target specifications, constraint rules, and a structured template into a single prompt. The LLM receives this prompt and produces two outputs: (i) initial estimates of process-dependent device parameters for every transistor ($\mu C_{ox}$, $\alpha g_m$, $\lambda$, $V_{th}$), informed by device constraints specified in the netlist, and (ii) a Python sizing function, `compute_sizing()`, encoding the topology's design equations. This task requires the LLM to interpret the raw netlist — recognizing the circuit topology, identifying device roles and signal paths, and deriving structurally correct design equations directly from the netlist. While the calibration feedback architecture is model-agnostic in principle, the framework's effectiveness depends on the LLM's netlist-interpretation capability. A deterministic execution step runs the sizing function with the LLM's parameter estimates, producing design variables ($W$, $L$, bias currents, etc). These are simulated in NGspice for the PTM models and in Cadence Virtuoso with the Spectre simulator for the TSMC PDK models (Sec. IV-C); the testbench and extracted metrics are described in Sec. IV-D. Finally, a calibration step, `calibrate()`, processes the simulation results: it extracts four calibration parameters per device from the DC OP, computes predicted-versus-actual performance, and assembles the Round 1 prompt.

TABLE I COMPARISON WITH EXISTING METHODS

| Method | LLM Role | Interpretable Equations | Pre-characterization | Cross-Node | Prediction-Error Analysis |
|---|---|---|---|---|---|
| BO [1] | None | No (black box) | None | Re-run required* | N/A |
| BO-Init [2] | None | No (black box) | Initial samples (16+) | Not demonstrated (45 nm only) | No |
| $g_m/I_D$ [5] | None | Partial (efficiency curves) | Lookup tables | Manual porting | N/A |
| EEsizer [6] | Optimizer | No | None | Yes (3 nodes) | No |
| LEDRO [7] | Search-space guide | No | None | Yes (4 nodes) | No |
| LLM-USO [8] | Knowledge transfer | No | None | Not specified | No |
| ADO-KT [9] | Knowledge transfer | No (knowledge graphs) | Prior-circuit runs | Not specified | No |
| AnaFlow [11] | Multi-agent | Partial (reasoning traces) | None | Not demonstrated (45 nm only) | No |
| Atelier [12] | Multi-agent + KB | Partial | Knowledge base | Yes (2 nodes) | No |
| AutoSizer [13] | Meta-optimizer | No | None | Not demonstrated (130 nm only) | No |
| TopoSizing [14] | Circuit understanding | Partial (annotations) | None | Not demonstrated (55 nm only) | No |
| EasySize [15] | Fine-tuned optimizer | No | Training data (350nm) | Yes (4 nodes)† | No |
| **This work** | **Equation derivation** | **Yes (Python code)** | **None (self-calibrating)** | **Yes (6 nodes)** | **Yes (183 rounds)** |

*'Re-run required' indicates the method must be re-executed from scratch at each new node.
†Trained on 350 nm; tested at 180 nm, 45 nm, and 22 nm nodes.
'This work' row summarizes the Table III benchmark (PTM, six nodes); the TSMC-PDK reproductions add runs at 40 nm and 180 nm (Sec. IV-C).

*2) Round N ≥ 1:* The LLM receives the prompt assembled by the previous round's calibration output and generates only an updated sizing function — it does not re-estimate device parameters, as `calibrate()` now supplies their simulation-extracted values. The updated function is executed, SPICE simulates, `calibrate()` extracts updated parameters, and `build_prompt()` assembles the next prompt. The loop completes when all specifications are met.

## *B. One-Shot Calibration*

In each round, the calibration extracts device-level parameters from a single DC OP simulation. No additional test structures, characterization sweeps, or lookup tables are required. For each transistor in the circuit, `calibrate()` computes four parameters.

*1) Effective transconductance parameter $\mu C_{ox}$ (A/V²):* Derived from the saturation current equation $I_d = ½\cdot\mu C_{ox}\cdot(W/L)\cdot V_{ov}^2$, where $V_{ov} = V_{gs} - V_{th}$ is the overdrive voltage. This captures the combined effect of carrier mobility, oxide capacitance, and mobility degradation at the actual operating point.

*2) Transconductance non-ideality factor $\alpha g_m$:* The ratio of actual $g_m$ to the ideal value $2\cdot I_d/V_{ov}$. This factor accounts for velocity saturation and short-channel effects that reduce $g_m$ below the long-channel prediction. In the tested processes, $\alpha g_m$ is typically below 1.0, depending on device type, channel length, and overdrive voltage.

*3) Channel-length modulation coefficient λ (1/V):* Computed as $g_{ds}/I_d$, where $g_{ds}$ is the output conductance from the OP. This parameter varies significantly with channel length and is critical for determining output resistance and voltage gain.

*4) Threshold voltage $V_{th}$ (V):* Extracted directly from the OP, including body-effect shifts for devices with non-grounded source terminals.

The one-shot extraction of these four parameters – together capturing the dominant process-specific effects – automatically adapts to the process node without modification to the calibration scripts. Since they depend on the device's operating point (which changes as *W*, *L*, and bias currents are adjusted each round), `calibrate()` re-extracts all four from every simulation.

The calibration data are presented to the LLM in a structured per-device table (Fig. 2) that includes the device type, *W/L*, measured current, overdrive voltage, transconductance, saturation status, and all four extracted parameters. The table also flags devices operating in the triode region ($|V_{ds}| < V_{ov}$), which alerts the LLM to bias-point failures requiring correction in the next round. Devices in triode or cutoff, or flagged for low overdrive, carry borrowed same-type saturation parameters, marked in the prompt.

The same calibration pass also extracts each device's operating-point parasitic capacitances ($C_{gs}$, $C_{gd}$, $C_{db}$, $C_{sb}$) from the simulator and supplies them to the model as a per-device table for pole and compensation estimates. On PTM BSIM3 models at 180 nm, which do not expose these quantities, the prompt directs the model to estimate device capacitances from *W*, *L*, and typical per-μm values instead (Table III note).

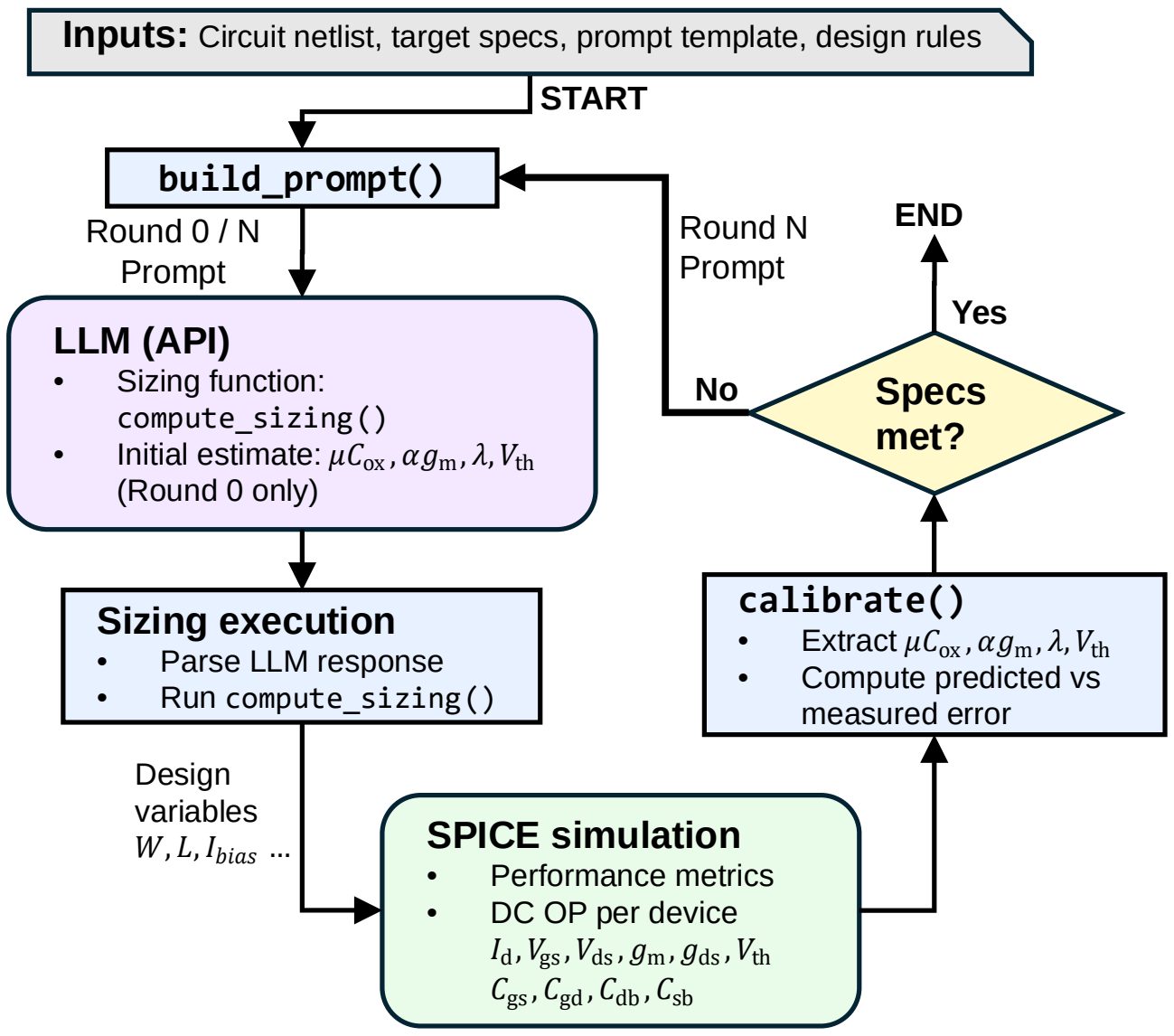

**Fig. 1.** Self-calibrating framework architecture. The loop iterates until all target specifications are met.

The operating-point dependence of these extracted quantities is substantial. Table II shows calibration data for the NMOS-input two-stage Miller-compensated cell (2SMC-N) at 45 nm – the same bench point as Fig. 2 – sampling the output and bias devices at rounds 0 and 2, two different points in the design space. The data illustrate three key observations.

First, the extracted parameters are operating-point quantities, not process constants: between rounds 0 and 2 the same devices' $\mu C_{ox}$ shifts from 1,748 to 1,146 μA/V² (M6) and from 361 to 204 μA/V² (M8) as their bias points move with the evolving design. Second, the extracted capacitances track the sizing itself: M8's $C_{gs}$ falls from 13.0 to 8.7 fF as its width is cut from 9.5 to 5.0 μm between the same rounds, giving the model fresh pole information each round. Third, $\alpha g_m$ remains well below unity throughout (0.31–0.64 in Table II), consistent with actual transconductance falling below the ideal long-channel value – precisely the correction the empirical $\alpha g_m$ supplies.

This calibration mirrors standard analog design practice: closed-form square-law reasoning structures the design around a verified DC operating point, and the simulator supplies everything the equations omit — mobility degradation, velocity saturation, short-channel and body effects, channel-length-modulation dependence, and nonlinear capacitances among them. The extracted parameters are accordingly local equivalents at the current operating point — used only to propose the next design, and never treated as global process constants.

## *C. Prediction-Error Feedback*

After each simulation round, `calibrate()` computes the discrepancy between the LLM's predicted performance and the simulator-measured results. The next-round prompt presents a side-by-side comparison and includes metric-type-specific

```
CALIBRATION FACTORS:
 Dev  Type     W/L         |Ids|   Vov     gm    Region    mu_Cox   a_gm     lam       ro    Vth
             (um/um)        (uA)  (mV)    (uS)            (uA/V2)          (1/V)    (kOhm)   (mV)
  M1  NMOS 43.95/0.15       7.74 -158.0  178.66  CUTOFF      0.0  0.000  0.0000      0.0  635.7
  M2  NMOS 43.95/0.15       7.76 -157.7  179.18  CUTOFF      0.0  0.000  0.0000      0.0  635.9
  M3  PMOS  2.71/0.20       7.74   61.2  120.24     SAT    304.4  0.476  0.1004   1287.5  585.2 *
  M4  PMOS  2.71/0.20       7.76   61.3  120.56     SAT    304.4  0.476  0.0995   1295.3  585.1 *
  M5  NMOS  2.29/0.20      15.50   53.7  196.69     SAT    935.3  0.341  6.6231      9.7  622.5 *
  M6  NMOS  0.57/0.20       5.00   55.2   73.68     SAT   1146.4  0.406  0.0833   2402.0  621.1 *
  M7  NMOS  4.59/0.20      39.57   54.9  584.14     SAT   1146.3  0.405  0.0868    291.2  621.4 *
  M8  PMOS  5.01/0.12      39.57   96.4  523.75     SAT    204.1  0.638  0.1991    127.0  581.9

  WARNING: M1 in CUTOFF (|Vgs|=478mV, Vov=-158mV)
  WARNING: M2 in CUTOFF (|Vgs|=478mV, Vov=-158mV)
  * |Vov| < 80 mV here: the extracted mu_Cox and a_gm are UNRELIABLE (square-law breaks down
    approaching weak inversion). When sizing FROM a flagged device, substitute a healthy
    same-type, same-L device that is well into saturation.

PARASITIC CAPACITANCES (from NGspice DC operating point, in fF):
  Dev  Type     Cgs(fF)  Cgd(fF)  Cdb(fF)  Csb(fF)
  --------------------------------------------------
    M1  NMOS     21.766   19.062    0.000    0.014
    M2  NMOS     21.771   19.463    0.000    0.014
    M3  PMOS      5.515    1.087    0.001    0.531
    M4  PMOS      5.518    1.052    0.001    0.531
    M5  NMOS      5.080    1.205    0.109    0.343
    M6  NMOS      1.277    0.227    0.000    0.080
    M7  NMOS     10.197    2.011    0.000    0.642
    M8  PMOS      8.658    2.261    0.002    0.397
```

**Fig. 2.** Per-device calibration factors and extracted capacitances from the round-2 prompt (2SMC-N, 45 nm). M1 and M2, flagged CUTOFF, carry borrowed same-type parameters (Sec. III-B).

margin formulas — proportional for linear metrics (e.g., GBW), additive for logarithmic metrics ($A_v$) and phase margin — embedded directly in the prompt text as formulaic instructions, so the LLM can compute corrected design targets arithmetically. A cumulative historical log of all prior rounds is included, enabling the LLM to identify persistent prediction biases and distinguish parametric errors, which shrink with calibration, from residual analytical-model limitations, which can persist across rounds.

The prediction-error feedback also includes targeted diagnostic guidance derived from failure modes observed during framework development. For example, a PM prediction exception instructs the LLM that if its predicted PM is catastrophically low but the measured PM passes, it should trust the measurement rather than distort the sizing to fix a prediction artifact. Such instructions encode domain knowledge about common analytical failure modes into the feedback loop; additional static constraints embedded in the prompt template are detailed in Section III-D.

This feedback is critical because certain analytical inaccuracies — such as systematic GBW over-prediction from parasitic capacitances whose effects the closed-form expressions only partially capture — are inherent in the analytical model, not in the extracted parameters, and persist after calibration. The feedback compensates by instructing the LLM to add design margin proportional to the observed over-prediction.

### *D. Structured Prompt Design*

The prompt enforces a multi-section output structure that requires the LLM to perform systematic topology analysis before writing any sizing code. The Round 0 prompt mandates seven sections in sequence:

(1) Device Roles — labeling each transistor's function (e.g., "NMOS diff-pair input," "PMOS diode-connected active load")
(2) Device Constraints — classifying each device
(3) Signal Path — tracing the signal from input to output
(4) Design Equations — deriving symbolic expressions for gain, GBW, power, and the dominant and non-dominant poles
(5) DC Bias Verification — checking that all branch voltages

TABLE II
CALIBRATED PARAMETERS AND EXTRACTED CAPACITANCES (2SMC-N, 45NM)

| Round | Dev | Type | $W/L$ (µm) | $V_{ov}$ (mV) | $\mu C_{ox}$ (µA/V²) | $\alpha g_m$ | $\lambda$ (1/V) | $V_{th}$ (mV) | $C_{gs}$ (fF) | $C_{gd}$ (fF) | $C_{db}$ (fF) | $C_{sb}$ (fF) |
|---|---|---|---|---|---|---|---|---|---|---|---|---|
| R0 | M6 | NMOS | 1.43/0.20 | 40.0 | 1748.4 | 0.312 | 0.087 | 621.1 | 2.83 | 0.57 | 0.000 | 0.190 |
| R0 | M8 | PMOS | 9.52/0.12 | 52.5 | 361.5 | 0.419 | 0.226 | 581.9 | 13.01 | 4.14 | 0.006 | 0.840 |
| R2 | M6 | NMOS | 0.57/0.20 | 55.2 | 1146.4 | 0.406 | 0.083 | 621.1 | 1.28 | 0.23 | 0.000 | 0.080 |
| R2 | M8 | PMOS | 5.01/0.12 | 96.4 | 204.1 | 0.638 | 0.199 | 581.9 | 8.66 | 2.26 | 0.002 | 0.397 |

Values extracted per round from the NGspice DC operating point.

support device saturation

(6) DC OP Consistency — verifying numerical compatibility of inter-stage node voltages

(7) Python Code with initial device estimates ($\mu C_{ox}$, $\alpha g_m$, $\lambda$, $V_{th}$)

This forced decomposition prevents the LLM from jumping directly to code generation without first reasoning about the circuit structure — a failure mode observed during early development that produced syntactically correct but topologically wrong sizing functions.

Within this structure, several specific constraints encode analog design practices that the LLM does not reliably apply unprompted.

*1) Device classification rules* require the LLM to label each transistor as MIRROR (sized by width ratio $W/W_{ref}$ relative to a reference device), MATCHED (identical $W$ and $L$ to a partner), or INDEPENDENT (sized directly by a performance equation). Each classification enforces a specific sizing formula in the Python code: mirror devices use $W = W_{ref} \times I_{target}/I_{ref} \times L/L_{ref}$, matched devices copy their partner's dimensions, and independent devices use the square-law equation $W = 2 \cdot I_d \cdot L\ /\ (\mu C_{ox} \cdot V_{ov}^2)$. Mirror devices must use the same channel length as their reference device, because $V_{th}$ depends on $L$ in short-channel processes. Violating this constraint was observed to cause mirror current errors exceeding 50%.

*2) An anti-circularity rule* requires the LLM to compute inter-stage node voltages strictly from the driving stage's device parameters (e.g., $V_{net5} = V_{dd} - V_{th,p} - V_{ov,8}$), preventing circular reasoning in which the LLM assumes a bias point and then verifies it against itself.

*3) DC OP consistency rules* require the LLM to show explicit numerical computation at every constrained node, verifying that the voltage one stage produces is compatible with what the next stage requires. The prompt requires explicit arithmetic at every headroom check, prohibiting the LLM from making vague assertions such as 'self-adjusts' or 'consistent by construction' without numerical verification.

*4) A parasitic capacitance warning* informs the LLM that large device widths create parasitic capacitance at internal nodes, degrading bandwidth and phase margin. The prompt instructs the LLM, when a high-frequency specification fails, to try reducing W (via higher overdrive voltage) rather than increasing it. This single prompt addition resolved a systematic convergence failure observed during development in which the LLM's response to failing PM was to increase current or device widths, inadvertently worsening the problem.

The LLM's output for each topology is a complete, commented Python function (`compute_sizing()`) that typically spans 200–300 lines. Fig. 3 shows a representative round-0 excerpt for a two-stage Miller-compensated OTA. Three features are notable: the LLM derives a joint GBW/SR constraint that selects the binding specification automatically, it applies all three device classification types (INDEPENDENT, MIRROR, MATCHED) with the correct sizing formula for each, and it cancels the RHP zero by setting $R = 1/g_{m8}$ — all without human guidance.

```python
# -- Current design: jointly set by GBW and SR --
# Relationship: SR_internal = 2*pi*GBW*Vov1
SR_internal_from_GBW = (
    2.0 * math.pi * GBW * Vov1)

if SR_internal_from_GBW >= SR_target:
    GBW_design = GBW              # GBW-limited
else:
    GBW_design = (                # SR-limited
        SR_target / (2.0 * math.pi * Vov1))

Cc = 0.5e-12
gm1_target = 2.0 * math.pi * GBW_design * Cc
Itail = gm1_target * Vov1
Id1 = Itail / 2.0

# Second-stage current: output SR and PM
Id2_SR = SR_target * CL
gm8_for_fp2 = (                  # fp2 >= 2.2*GBW
    2.2 * GBW * 2.0 * math.pi * CL)
Id2_fp2 = gm8_for_fp2 * Vov8 / 2.0
Id2 = max(Id2_SR, Id2_fp2)

# -- Device widths --
# M6: INDEPENDENT (PMOS ref, carries Iref)
W6 = 2.0*Iref*L6 / (mu_Cox_p * Vov6**2)

# M5: MIRROR of M6, carries Itail
W5 = W6 * (Itail / Iref) * (L5 / L6)

# M7: MIRROR of M6, carries Id2
W7 = W6 * (Id2 / Iref) * (L7 / L6)

# M1, M2: MATCHED diff pair (each carries Id1)
W1 = 2.0*Id1*L1 / (mu_Cox_p * Vov1**2)
W2 = W1

# M8: INDEPENDENT (NMOS CS gain, carries Id2)
W8 = 2.0*Id2*L8 / (mu_Cox_n * Vov8**2)

# -- Compensation (nulling RHP zero) --
gm8_sized, Vov8_v = calc_gm("M8", Id2, W8, L8)
R = 1.0 / gm8_sized
```

**Fig. 3.** Excerpt from an LLM-generated `compute_sizing()` function for a two-stage Miller-compensated OTA, Round 0. The complete function (~200 lines) is executed unmodified.

The framework requires no topology-specific prompt tuning: the same prompt template, unchanged, serves every circuit from the 6-transistor single-stage OTA to the 30-transistor class-AB opamp, and every process node from 32 nm to 180 nm. Beyond the per-run problem statement (netlist, target specifications, and process node), nothing in it names a circuit or supplies circuit-specific equations; the device classifications it requests — mirror, matched, independent — are outputs the model must derive from the netlist, not information supplied to it.

## IV. Experimental Setup

### A. Technology and Specifications

The framework is evaluated at six process nodes (32 nm to 180 nm), using the PTM models [18] with the following per-node supplies:

- *32 nm CMOS:* $V_{dd}$ = +0.5 V, $V_{ss}$ = −0.5 V.
- *45 nm CMOS:* $V_{dd}$ = +0.55 V, $V_{ss}$ = −0.55 V.
- *65 nm CMOS:* $V_{dd}$ = +0.55 V, $V_{ss}$ = −0.55 V.
- *90 nm CMOS:* $V_{dd}$ = +0.6 V, $V_{ss}$ = −0.6 V.
- *130 nm CMOS:* $V_{dd}$ = +0.65 V, $V_{ss}$ = −0.65 V.
- *180 nm CMOS:* $V_{dd}$ = +0.9 V, $V_{ss}$ = −0.9 V.

The Cadence reproductions use the TSMC PDKs at 40 nm and 180 nm with their PDK-specified supplies (Sec. IV-C). The evaluation comprises two sets of experiments:

*1) OTA validation:* The first set validates generality across OTA topologies and process nodes. The six benchmark families of Sec. IV-B are tested one node per cell across 32-180 nm, with per-cell load capacitances and target specifications given in Table III. Targets are set per cell, reflecting both the topology's capabilities and the node's supply: gain targets follow the topology, from 30-35 dB for the single-stage cells up to 70 dB for the most demanding cells; output-swing and power budgets scale with the per-node supply (0.60-1.10 $V_{pp}$ and 0.08-0.45 μW across the ladder); and speed targets are chosen per cell within each node's capability rather than by a single scaling rule.

*2) Matched-specification comparisons:* The second set validates the framework against four published methods — James et al. [2], LLM-USO [8], Atelier [12], and EEsizer [6] — on matched circuits and specifications, each comparison using five independent trials with targets, load capacitance, and load resistance matched to the reference work (methodology and per-work targets in Section IV-C).

## *B. Topologies Tested*

Six OTA topology families are tested, each in NMOS- and PMOS-input variants with one process node per cell — twelve cells in total (Table III), plus two comparison anchors from Sec. IV-C that complete the circuit set. All circuits are differential-input, single-ended-output designs, internally biased through current mirrors driven by a single ideal reference current source ($I_{ref}$); ideal passive components belonging to a topology — compensation capacitors and nulling resistors — are noted with the circuits below. Items 1-6, the benchmark families of Table III, appear in order of transistor count, followed by the two comparison anchors:

*1) 1S-N, 1S-P* (6 transistors; Fig. 4(a)): single-stage OTA.

*2) 2SMC-N, 2SMC-P* (8 transistors; Fig. 4(b)): a two-stage Miller-compensated OTA with $C_c$ and $R_z$.

*3) CM-N, CM-P* (10 transistors; Fig. 4(c)): a current-mirror OTA.

*4) FC-N, FC-P* (18 transistors; Fig. 4(e)): a folded-cascode OTA.

*5) CMC-N, CMC-P* (18 transistors; Fig. 4(f)): a current-mirror OTA with a cascoded output stage for high output impedance.

*6) GBFC-N, GBFC-P* (28 transistors; Fig. 4(d)): a gain-boosted folded cascode in which two auxiliary amplifiers regulate the output cascodes.

*7) NMC* (16 transistors; Fig. 4(g)): a three-stage nested-Miller-compensated amplifier with two nested RC compensation branches ($R_1$-$C_1$, $R_2$-$C_2$); the Atelier comparison target (Sec. IV-C).

*8) 30T* (30 transistors; Fig. 4(h)): the 30-transistor rail-to-rail input/output (RRIO) class-AB amplifier with two Miller compensation capacitors ($C_{C1}$,$C_{C2}$); the EEsizer comparison target (Sec. IV-C).

To ensure results do not rest solely on recall of standard formulas, the set includes circuits that are not the typical textbook-covered topologies — the gain-boost arrangement of item 6 and the two anchors of items 7 and 8 — complemented by the disguised-netlist controls of Sec. V-F.

Fig. 4 shows the NMOS-input variant of each benchmark family; their PMOS-input counterparts in Table III are equivalent, with complementary device types and mirrored polarity. Testing six topology families in both input polarities across six process nodes, plus the NMC and 30T anchors, provides fourteen independent validation points for the framework's generality across circuit complexity, compensation architectures, and process nodes; four of these cells double as the comparison circuits of Sec. IV-C.

The framework's purpose is the reliable sizing of known topologies to given requirements, not the invention of new ones; it therefore relies on the model's ability to reason from the netlist and to recognize the structure it is sizing, and it inherits a dependence on the capability of the underlying model — a dependence we quantify across five models in Sec. V-E (Table V).

## *C. Matched-Specifications Comparison Methodology*

To enable controlled comparison with published methods, we apply the framework to circuits and specification targets matching those reported in four recent works, noting differences in process technology.

*1) James et al. [2]:* 2SMC-N and FC-N at 45 nm, targeting $A_v \geq 45$ dB, GBW ≥ 20 MHz, PM ≥ 45°, Power ≤ 200 μW. We reproduce their targets at 45 nm PTM in NGspice and, in Cadence Virtuoso, at 40 nm TSMC — the closest PDK available to us (Table IV). Load capacitance was not specified in [2]; we use $C_L$ = 0.5 pF as a conservative estimate of parasitic loading.

*2) LLM-USO [8]:* FC-N at 90 nm PTM, targeting $A_v \geq 60$ dB, unity-gain frequency (UGF) ≥ 1 MHz, common-mode rejection ratio (CMRR) ≥ 80 dB, $C_L$ = 1 pF. The process node is not specified in [8]; we adopt 90 nm PTM while matching their specification targets.

*3) Atelier [12]:* NMC at 180 nm, targeting $A_v > 85$ dB, GBW > 0.7 MHz, PM > 55°, Power < 50 μW, input offset (Offset) < 3 mV, input-referred noise (Noise) < 500 nV/rtHz at 1 kHz, transistor area (Area) < 250 μm$^2$, $C_L$ = 10 pF, $R_L$ = 1 MΩ. We reproduce their targets at 180 nm on both PTM in NGspice and the TSMC PDK in Cadence Virtuoso (Table IV); Atelier's full pipeline also spans topology selection, so the comparison here is on its sizing stage.

*4) EEsizer [6]*: 30T at 180 nm PTM, targeting $A_v \geq 65$ dB, UGBW ≥ 10 MHz, PM ≥ 50°, Power ≤ 10 mW, CMRR ≥ 100 dB, total harmonic distortion (THD) ≤ -26 dB, input-referred offset (Offset) ≤ 1 mV, Output Swing ≥ 1.53 V, input common-mode range (ICMR) ≥ 1.71 V, $C_L$ = 10 pF, $R_L$ = 1 kΩ. EEsizer specifies swing and ICMR as $V_{dd} - V_{ss}$ (1.8 V) with a 5% tolerance applied to all metrics; we adopt the same effective threshold, 0.95 × ($V_{dd} - V_{ss}$) = 1.71 V, as the swing and ICMR targets. Matching their node study, we additionally run the cell at 130 and 90 nm PTM, carrying all targets unchanged except swing and ICMR, which follow the same rule at each node's supply (1.23 V and 1.14 V). EEsizer validates a 20-transistor version of this topology with a manually designed bias circuit; our implementation includes all 30 transistors as design variables.

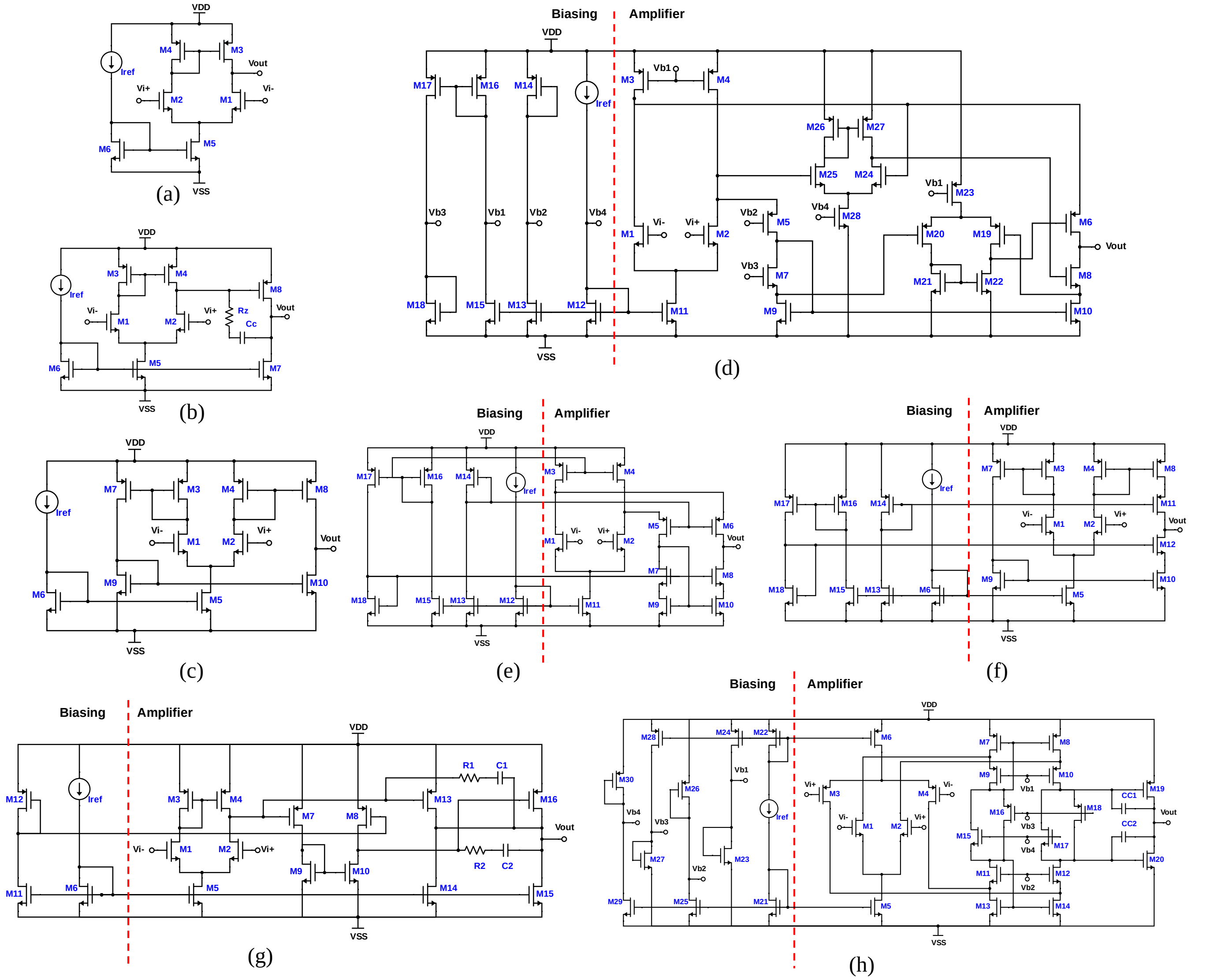

**Fig. 4.** Benchmark and comparison-anchor topologies (NMOS-input variants shown). (a) Single-stage (1S). (b) Two-stage Miller-compensated (2SMC). (c) Current-mirror (CM). (d) Gain-boosted folded cascode (GBFC). (e) Folded-cascode (FC). (f) Cascoded current-mirror (CMC). (g) Nested-Miller-compensated (NMC), the Atelier target. (h) 30-transistor class-AB (30T), the EEsizer target.

### *D. Experimental Controls*

The following controls ensure reproducibility and prevent data contamination: Each round is issued as an independent API call with no memory of prior rounds — every round's prompt is the model's entire context (Sec. VI-B). Each topology is tested independently with no cross-topology learning. The primary LLM is Claude Opus 5 (Anthropic); five LLMs in total are evaluated on a shared cross-model bench (Sec. V-E). Only the structured prompt and calibration data are provided — no per-topology design hints, manual parameter adjustments, or iterative coaching. All `compute_sizing` code generated by the LLM is executed unmodified.

Two execution flows are used: in the NGspice flow the model is driven through its API, and in the Cadence reproductions through its web interface, a fresh chat session per round (Sec. IV-C). The testbench is identical across all topologies and both execution flows, configured in unity-gain feedback. The low-frequency gain ($A_v$) at 1 Hz, GBW, and PM are extracted from the loop-gain (STB) analysis, while SR is measured from a transient large-signal step response. CMRR is obtained from the ratio of differential to common-mode gain in AC at 1 Hz, with nominal DC voltage set to 0 V. Input-referred noise is the spot noise density at 1 kHz. Power and input-referred offset are computed from the DC OP, and THD is measured from a transient simulation with a 100 kHz sinusoidal input sized to drive the output to 80% of its measured linear swing. Finally, ICMR and

TABLE III CONVERGENCE RESULTS (12-CELL PTM BENCHMARK)

| Node | 32nm | | 45nm | | 65nm | | 90nm | | 130nm | | 180nm | |
|---|---|---|---|---|---|---|---|---|---|---|---|---|
| Circuit | 1S-N | 2SMC-P | 2SMC-N | CM-N | CMC-N | GBFC-P | 1S-P | GBFC-N | CMC-P | FC-N | CM-P | FC-P |
| CL (pF) | 0.075 | 0.075 | 0.1 | 0.1 | 0.15 | 0.15 | 0.25 | 0.25 | 0.5 | 0.5 | 1 | 1 |
| Conv | 5/5 | 5/5 | 5/5 | 5/5 | 5/5 | 5/5 | 5/5 | 5/5 | 5/5 | 5/5 | 5/5 | 5/5 |
| Rounds | 2.6±0.9 | 3.6±0.9 | 6.4±2.3 | 3.2±0.8 | 2.0±0.0 | 7.4±3.0 | 1.2±0.4 | 1.8±0.8 | 1.6±1.3 | 1.6±0.5 | 3.2±1.3 | 2.0±0.7 |
| Time (min) | 30±16 | 58±19 | 181±102 | 42±14 | 28±7 | 200±95 | 10±6 | 23±9 | 19±19 | 17±6 | 43±17 | 19±7 |
| Av (dB) | 41.1/30 | 86.9/70 | 52.7/50 | 44.6/40 | 71.5/60 | 73.5/70 | 43/35 | 86.3/70 | 64.5/60 | 63/50 | 40.7/40 | 61.8/60 |
| GBW (MHz) | 64/60 | 63/50 | 411/200 | 161/100 | 130/100 | 164/80 | 182/90 | 128/85 | 54/50 | 60/45 | 74/55 | 106/100 |
| PM (deg) | 116/60 | 65.6/45 | 64.2/60 | 76.3/60 | 64.7/60 | 93.9/60 | 92.6/60 | 61/60 | 82/60 | 77.1/60 | 47.4/45 | 72/60 |
| Power (μW)† | 6.4/80 | 60.4/80 | 96/100 | 31.7/100 | 39/100 | 84.1/100 | 61/120 | 89.2/120 | 102/130 | 95.7/130 | 349/450 | 425/450 |
| CMRR (dB) | 53.5/35 | 47.4/35 | 49.3/35 | 64.8/35 | 94.1/35 | 75.9/65 | 69.3/50 | 74.7/55 | 73.7/40 | 79.9/70 | 80.5/50 | 88/60 |
| Noise (nV/rtHz)† | 88/300 | 226/400 | 186/400 | 339/500 | 353/600 | 111/300 | 208/400 | 249/450 | 363/450 | 161/300 | 8.7/250 | 9.3/250 |
| THD (dB)† | -14/-10 | -16/-12 | -20/-10 | -18/-10 | -39/-30 | -67/-40 | -37/-30 | -38/-35 | -29/-25 | -57/-45 | -21/-15 | -55/-40 |
| SR+ (V/us) | 19/1 | 86/30 | 349/100 | 79/15 | 66/25 | 32/30 | 69/40 | 93/35 | 29/20 | 34/25 | 32/30 | 61/50 |
| SR- (V/us) | 2.4/1 | 135/30 | 106/100 | 29/15 | 57/25 | 98/30 | 156/40 | 41/35 | 44/20 | 35/25 | 197/30 | 77/50 |
| ICMR (V) | 0.78/0.2 | 0.71/0.2 | 0.87/0.2 | 0.87/0.2 | 0.84/0.3 | 0.72/0.3 | 0.96/0.3 | 0.89/0.3 | 1.12/0.3 | 0.80/0.3 | 1.50/0.4 | 1.02/0.4 |
| Swing ($V_{pp}$) | 0.76/0.6 | 0.84/0.7 | 0.89/0.75 | 0.86/0.7 | 0.84/0.7 | 0.71/0.7 | 0.95/0.8 | 0.94/0.8 | 1.15/0.8 | 0.79/0.75 | 1.49/1.1 | 1.01/0.9 |
| Offset (mV)† | 0.76/5 | 0.29/2 | 0.51/5 | 1.37/5 | 2.15/5 | 0.33/2 | 1.00/5 | 0.28/5 | 1.21/5 | 0.17/5 | 1.21/5 | 0.15/5 |

Specification cells show measured/target from each cell's median-convergence trial (chosen to avoid selection bias). Conv, Rounds, and Time rows are mean±std across the five trials. The 180 nm columns (PTM BSIM3) lack per-device capacitance quantities; the Sec. III-B proxy fallback applies.
†Lower-is-better metrics (measured ≤ target).

output swing are measured using a DC sweep, identifying the input and output voltage ranges over which $dV_{out}/dV_{in} \geq 0.9$.

Convergence is defined as simultaneously meeting all target specifications in a single simulation. Simulation counts include Round 0 (the initial sizing from LLM estimates); thus, a topology converging at Round N requires N + 1 total simulations.

Beyond the benchmark itself, controlled arms complement it: ablations D1-D5 that remove framework components individually (Sec. V-E); obfuscated-netlist controls with permuted device indices and neutral net names (Sec. V-F); parameter-perturbation and guidance-removal probes; and a classical Bayesian-optimization baseline (TuRBO-1) run on identical benches. Benchmark and comparison conditions use N = 5 independent trials; the ablation and control studies use N = 3 (N = 5 where noted), with trial counts stated in each table.

## V. RESULTS

Sections V-A and V-B present convergence results and prediction-error analysis for the twelve benchmark cells, and Sec. V-C dissects the hardest cell. Sec. V-D presents the matched-specification comparisons against James et al. [2], LLM-USO [8], Atelier [12], and EEsizer [6]. Sec. V-E reports the cross-model and ablation studies, and Sec. V-F the verification and robustness checks.

### A. Convergence Summary

Table III reports the full benchmark: all twelve cells converge in every trial — 60 of 60 runs with the primary model — with mean rounds per cell ranging from 1.2 to 7.4; the specification values shown are those of each cell's median-convergence trial (the middle trial by rounds, so the displayed design is representative rather than best-case). Convergence difficulty scales with structural complexity: the gain-boosted folded cascode at 65 nm is the slowest cell (7.4 ± 3.0 rounds), examined in Sec. V-C, while the single-stage and current-mirror cells typically close in one to three rounds. Every entry in Table III rests on five independent trials per cell.

### B. Prediction Accuracy and Cross-Node Comparison

Across the twelve benchmark cells (Table III), initial prediction accuracy is only a weak guide to convergence. Cells with tighter design windows tend both to start with larger errors and to need more rounds, but the extremes break any direct mapping: the largest round-0 GBW error in the benchmark (the single-stage OTA at 32 nm, median 2,776%) converges in 2.6 rounds; the slowest cell (GBFC-P, 7.4 rounds) starts at only 23%; and the most accurately predicted cell (GBFC-N, median 2%) converges quickly (1.8 rounds), yet three cells with larger errors converge faster. Beyond the extremes, the agreement is only partial: taking GBW as the test case, ranking the cells by its round-0 error and by their mean rounds gives Spearman $\rho$ = 0.62, and no specification's round-0 error ranks the cells substantially better (the strongest, noise, reaches 0.69). The width of the design window, not prediction accuracy, sets the pace: Sec. V-C develops this feasibility argument on the hardest cell, and the matched-specification comparisons of Sec. V-D show the same behavior on additional circuits.

Resolving prediction error by round shows where accuracy comes from. Before any measurement, the model's initial predictions are unreliable: the mean GBW error at round 0 is 499%, dominated by a small number of far-off estimates, though the median is 23%. One calibrated feedback round changes the picture — the mean drops to 33% and the median to 20% — and by round 3 the median reaches 10%. Phase-margin prediction improves in the same way once feedback begins: the median PM error falls from 14.8° at round 0 to 12.9° after one round and 9.3° by round 3. Slew-rate prediction follows the same pattern from larger initial errors: medians of 76% (SR+) and 39% (SR−) at round 0 fall to roughly 21% and 31% after the first feedback round. The remaining specifications follow the same round-over-round correction.

The ablation arms (Sec. V-E) separate where these gains come from. Removing the calibrated operating-point data (arm D1) costs accuracy: over the feedback rounds, the median GBW error more than doubles relative to the matched full-framework runs (38.5% versus 18.5%) — yet every D1 trial in Table VI still converges, because the loop still steers on measured performance. Removing the feedback history (D2) or the structured methodology (D3) leaves prediction accuracy intact but costs reliability, with failed trials on the hardest cells; removing the loop entirely (D5) prevents convergence altogether on the cells tested (Table VI). Together, the arms separate the roles: measured feedback produces convergence, calibration makes the predictions accurate, and the history and methodology protect reliability on the hardest cells.

### *C. Difficulty Anatomy: The GBFC-P Cell*

The gain-boosted folded cascode at 65 nm is the benchmark's hardest cell: 7.4 ± 3.0 rounds at baseline versus a median of two rounds across the other cells, the only cell where ablations cost Opus 5 reliability (Sec. V-E), and a cell GPT-5.6 cannot close (0/5, best 11 of 12; Table V). The difficulty is structural: the two auxiliary amplifiers couple the output cascodes' bias to the main signal path, so sizing the boost loops interacts with gain, swing, and stability simultaneously, and errors surface as long, multi-round corrections rather than as isolated specification misses. Under ablation, these corrections escalate into the response-length failures noted in Sec. V-E. The design trail makes this visible: revisions concentrate on the auxiliary amplifiers and the output branch, exactly where the coupling lives.

### *D. Comparison with Prior Methods*

The comparisons with prior methods are organized into matched-specification blocks — summarized in Table IV and reported specification by specification in Supplementary Table S1 — built under the methodology of Sec. IV-C: for each anchor, the circuit, targets, and results are taken directly from that work's published paper; our runs reproduce the same targets at the same or the closest realizable node; and every residual difference — circuits, loads, supplies, success criteria, trial counts, measurement conventions — is stated in the table notes. Entries the anchor papers do not report are left as dashes, and no cross-work superiority in simulation count is claimed: conditions across works are aligned, not identical — models, interfaces, and some conditions necessarily differ, so even where our counts are lower the comparison cannot be read as one-to-one — and the blocks are offered as indicative alignment under matched targets. The framework's central contribution is not the count itself but a system that sizes accurately, converges reliably, and leaves an explicit design trail each round through its `compute_sizing` function.

Two further steps tighten the comparison, one bridging the process models and one adding a fully matched classical baseline. In the effort to replicate the anchors' testing environments where their exact fabrication processes were unavailable, the same targets are run on both model sets: for Atelier, the nested-Miller cell on the TSMC 180 nm PDK in Cadence Virtuoso and on PTM 180 nm in NGspice; for James, the two cells at 45 nm on PTM in NGspice and on the closest available TSMC PDK, 40 nm, in Cadence Virtuoso — demonstrating each comparison on both an industrial PDK and the open PTM models, so the results do not hinge on either process model (TSMC rows, Table IV). The EEsizer node runs use the same PTM model family with per-node supplies. As the fully matched baseline, TuRBO-1, a strong classical Bayesian-optimization method, is re-run on the identical PTM netlists, testbenches, specifications, and success criteria as our runs (Sec. IV-D; TuRBO-1 row, Table V). Where a method does not reach the targets within its simulation budget, the table reports the budget rather than a simulations-to-success value.

TABLE IV CONVERGENCE COMPARISON SUMMARY

| Method | Circuit | #T | Node (nm) | Sims u±s | Succ. |
|---|---|---|---|---|---|
| James [2] | 2SMC-N | 8 | 45 TSMC | 21-28† | 1/1 |
| This work | | | 45 PTM | 2.0 ± 0.0 | 5/5 |
| | | | 40 TSMC* | 1.7 ± 0.6 | 3/3 |
| James [2] | FC-N | 13 | 45 TSMC | 21-28† | 1/1 |
| This work | | 18 | 45 PTM | 2.0 ± 0.0 | 5/5 |
| | | | 40 TSMC* | 2.3 ± 0.6 | 3/3 |
| LLM-USO [8] | FC-N | 13 | 90‡ | 45 | 1/1 |
| This work | | 18 | 90 PTM | 1.0 ± 0.0 | 5/5 |
| Atelier [12] | NMC | 13-16§ | 180 SMIC | 100-400 | 10/10 |
| This work | | 16 | 180 PTM | 1.2 ± 0.4 | 5/5 |
| | | | 180 TSMC* | 2.0 ± 0.0 | 3/3 |
| EEsizer [6] | 30T | 20 | 180 PTM | 13/15 | 2/5 |
| This work | | 30 | | 1.4 ± 0.5 | 5/5 |
| EEsizer [6] | 30T | 20 | 130 PTM | 22 | 1/5 |
| This work | | 30 | | 2.4 ± 0.9 | 5/5 |
| EEsizer [6] | 30T | 20 | 90 PTM | - | 0/5 |
| This work | | 30 | | 2.2 ± 1.3 | 5/5 |

*Cadence Spectre, web-interface flow (Sec. IV-D).
†Range across circuits and initialization strategies; $C_L$ not specified in [2], 0.5 pF here.
‡Node not stated in [8]; inferred from the minimum channel length.
§Amplifier core is 13 transistors; up to 3 additional bias transistors not shown in [12].
[6] evaluates a 20T op-amp with externally fixed bias; ours is a 30T variant biased from a single current reference (non-identical netlists; identical loads).
[6]'s Sims cells list the successful trials' iteration counts; at 90 nm all five attempts exhausted the 25-iteration budget. Its supplies differ from ours only at 130 nm ([6] retains 1.8 V; this work 1.3 V). Some measurement methods differ between works (Sec. IV-D).

Table IV summarizes the comparisons by circuit family: against James et al. [2], the two 4-specification cells close at matched targets in one to a few rounds, with the TSMC rows carrying the same targets onto the foundry PDK; against LLM-USO [8], the folded cascode closes its three matched targets at 90 nm; against Atelier [12], the nested-Miller amplifier closes on both PTM and the TSMC bridge; and against EEsizer [6], the 30-transistor amplifier closes on the same PTM family at 180 nm and at the scaled 130 nm and 90 nm nodes. The methodological contrast is qualitative rather than numerical: prior methods return final sizes or logged conversations, whereas this framework leaves a per-round executable rationale for every device dimension (Sec. V-F).

### *E. Cross-Model and Ablation Studies*

Table V compares five models running the same framework on a shared six-cell bench: the 2SMC-N and FC-N cells at 45 nm replicating the specifications of James et al. [2] (Sec. IV-C), three benchmark cells (1S-P, CM-N, and GBFC-P), and the 30T amplifier replicating EEsizer [6]. A comparison to TuRBO-1 is also included (five independently seeded runs per cell; simulations-to-success under a 600-evaluation budget, 1,000 for the 30T; Sec. V-D). Across this bench, cross-model reliability falls as circuit difficulty increases, and the two hardest circuits separate the tiers: the 30T is closed by the top three models and by neither o3 nor Haiku, while the gain-boosted folded cascode is closed only by Opus 5 (per-cell counts and rounds in Table V). Across models, the unconverged runs end as designs still short of one or more specifications when the round budget is reached — the sizing code executes and the simulations run; however, the designs do not converge. On the gain-boost cell, o3's revisions are locally coherent, correctly diagnosing bias collapses and prescribing sensible fixes; the round trail nonetheless shows them trading one specification cluster against another — recovering swing and input range at the cost of offset and CMRR and vice versa — without reaching a jointly feasible point. Every converged run's equations pass the structural audit of Sec. V-F; the model dependence therefore localizes in the capacity for sustained, multi-constraint revision, not in netlist understanding.

Complementing the cross-model view, Table VI isolates the framework's contribution from the model's: five ablation arms, run with both GPT-5.6 and Claude Opus 5, each remove a defined set of components from the prompts. D1 removes the measured operating-point data and everything derived from it, retaining history and the prediction-vs-actual comparison; D2 removes the feedback history, the prediction-vs-actual comparison, and the error-margin guidance, retaining all measured and calibrated data; D3 removes the structured design methodology (the seven-section analysis and the sizing-method rules, replaced by a bare estimate-and-code request) together with the history, comparison, and margin guidance, retaining all measured and calibrated data; D4 combines all three removals, leaving the netlist, the targets, and raw measured performance only; and D5 removes the LLM entirely, solving a fixed, hand-written analytical equation set each round against the newly measured calibration data. All arms retain the minimum interface — the netlist, target specifications, design-variable list, the previous round's `compute_sizing`, the required output format, and raw measured results — without which the loop could not run.

For GPT-5.6, removals produce failures on cells it otherwise closes 5/5: the 30T amplifier drops to 2/3 converged when the feedback history, the methodology, or all three layers are removed (D2–D4), and the FC-N cell slows from 3.8 to 6.7 mean rounds without the measured operating-point data (D1). For Opus 5, the simpler cells still converge under every removal, but the separation emerges as difficulty rises: on the hardest cell (GBFC-P), the history and methodology removals cost reliability (3/5 and 4/5 converged, with response-length failures, versus 5/5 at baseline) and the bare loop costs efficiency (11.0 versus 7.4 mean rounds), while the calibration removal converges every trial there and pays instead in prediction accuracy (Sec. V-B). Performance therefore degrades even with Opus 5; the full framework nonetheless retains a distinct contribution in every configuration: each feedback round's `compute_sizing` constitutes an explicit, auditable design rationale, grounded in the measured device data and revised against the recorded prediction errors — the design trail examined in Sec. V-F. Every ablated arm forfeits at least one of the two. Finally, the model itself is indispensable: in D5 — the fixed equation set re-solved each round with no LLM in the loop — no trial ever closes over the full 30-round budget; the best reaches 10 of 12 specifications on the easiest cell (3 of 12 on 2SMC-N), and even that peak is not held — it regresses to 7 of 12 as the calibration data shifts under the fixed equations. Fixed equations plus live calibration do not converge; closing the gap would require a designer revising the equations after every round — and each such revision is a design act: re-choosing overdrives, re-budgeting bias currents, re-working the compensation as the measured data shift — which is precisely the step the LLM performs in the loop.

TABLE V CROSS-MODEL EVALUATION

| Model | 1S-P 90nm | CM-N 45nm | GBFC-P 65nm | 2SMC-N 45nm | FC-N 45nm | 30T 180nm |
|---|---|---|---|---|---|---|
| Opus 5 | 5/5<br>1.2±0.4 | 5/5<br>3.2±0.8 | 5/5<br>7.4±3.0 | 5/5<br>2.0±0.0 | 5/5<br>2.0±0.0 | 5/5<br>1.4±0.5 |
| GPT-5.6 | 5/5<br>1.2±0.4 | 4/5<br>8.0±1.6* | 0/5<br>[11/12] | 5/5<br>2.4±0.5 | 5/5<br>3.8±2.7 | 5/5<br>3.0±0.7 |
| Opus 4.8 | 5/5<br>1.8±1.3 | 0/5<br>[11/12] | 0/3<br>[10/12] | 5/5<br>2.0±0.0 | 5/5<br>2.6±0.9 | 5/5<br>7.4±3.4 |
| o3 | 3/5<br>17.3±10.7* | 2/5<br>14.5±3.5* | 0/5<br>[8/12] | 5/5<br>5.6±2.9 | 4/5<br>14.2±3.1* | 0/5<br>[8/9] |
| Haiku 4.5 | 3/5<br>9.0±7.9* | 0/5<br>[10/12] | 0/5<br>[8/12] | 5/5<br>10.2±3.6 | 0/5<br>[3/4] | 0/5<br>[8/9] |
| TuRBO-1 | - | - | 0/5 | 5/5<br>42.4±36.2 | 5/5<br>75.6±30.0 | 0/5 |

Entries: converged/attempted; second line: rounds mean ± std; brackets: best specifications reached when no trial converged.
TuRBO-1's second line counts simulations, not rounds; its unconverged cells carry no second line.
*Over the converged trials only.

TABLE VI ABLATION STUDY

| Cell (model) | Full | D1 | D2 | D3 | D4 |
|---|---|---|---|---|---|
| 1S-P 90nm<br>Opus 5 | 5/5<br>1.2±0.4 | 3/3<br>1.7±0.6 | 3/3<br>1.0±0.0 | 3/3<br>1.7±0.6 | 3/3<br>1.0±0.0 |
| 1S-P 90nm<br>GPT-5.6 | 5/5<br>1.2±0.4 | 3/3<br>1.3±0.6 | 3/3<br>1.0±0.0 | 3/3<br>1.0±0.0 | 3/3<br>1.7±1.2 |
| 2SMC-N 45nm<br>Opus 5 | 5/5<br>2.0±0.0 | 3/3<br>2.0±0.0 | 3/3<br>2.0±0.0 | 3/3<br>2.0±0.0 | 3/3<br>1.0±0.0 |
| 2SMC-N 45nm<br>GPT-5.6 | 5/5<br>2.4±0.5 | 3/3<br>3.3±0.6 | 3/3<br>2.3±0.6 | 3/3<br>2.0±0.0 | 3/3<br>1.3±0.6 |
| FC-N 45nm<br>Opus 5 | 5/5<br>2.0±0.0 | 3/3<br>2.3±0.6 | 3/3<br>2.0±0.0 | 3/3<br>1.7±0.6 | 3/3<br>2.7±1.2 |
| FC-N 45nm<br>GPT-5.6 | 5/5<br>3.8±2.7 | 3/3<br>6.7±4.2 | 3/3<br>2.0±0.0 | 3/3<br>2.7±1.2 | 3/3<br>1.7±0.6 |
| 30T 180nm<br>Opus 5 | 5/5<br>1.4±0.5 | 3/3<br>2.3±0.6 | 3/3<br>1.3±0.6 | 3/3<br>2.0±0.0 | 3/3<br>3.0±0.0 |
| 30T 180nm<br>GPT-5.6 | 5/5<br>3.0±0.7 | 3/3<br>3.0±1.7 | 2/3<br>3.5±3.5* | 2/3<br>3.5±2.1* | 2/3<br>5.0±1.4* |
| GBFC-P 65nm<br>Opus 5 | 5/5<br>7.4±3.0 | 5/5<br>4.0±1.2 | 3/5<br>6.0±3.0* | 4/5<br>6.5±3.7* | 5/5<br>11.0±3.6 |

Entries: converged/attempted; the second line: rounds mean ± std.
Full: the matched full-framework baselines (Secs. V-A, V-D).
D1: no calibrated operating-point data; D2: no history or prediction comparison; D3: no design methodology (nor history/comparison); D4: all three removed (bare loop); definitions in Sec. V-E.
D5 (fixed hand-written equations, no LLM; model-independent) never converges; results in the text.
*Over the converged trials only.

A third study addresses sensitivity to errors in the calibration data itself: on the 2SMC-N and FC-N cells (GPT-5.6, three trials per level), the calibration factors are corrupted by ±5, ±10, and ±20% before each round. Every trial converges at every level, and mean rounds at ±20% are 2.7 on both cells, against matched baselines of 2.4 and 3.8. The loop compensates because each round's prompt also carries the unmodified measured operating-point data and the measured performance itself, so the model corrects against ground truth (per-level accuracy in Supplementary Table S2).

### *F. Verification and Robustness*

Structural correctness — the requirement that the LLM correctly identifies which devices govern each specification and the direction of their influence — is central to the convergence claim. Table VII shows an exemplar: the final-round equations of the 2SMC-N median-convergence trial — the same design whose measured values Table III reports — each with its governing devices and the direction of its influence. The equations carry the canonical structure: the two-stage gain decomposition, GBW set by the first-stage transconductance over $C_c$, and the Miller-specific slewing split between the tail charging $C_c$ and the output stage discharging the load. Their numerical anchors — the transconductances, mirror ratios, node capacitances, and the SR− tail-collapse factor — come from the preceding rounds' measured operating points, and two predictions (THD, offset) are held at their measured values.

Table VII's rows are audited individually in Supplementary Table S6; beyond the exemplar, the audit covers everything the framework produced. An AST-based extractor read the final-round `compute_sizing` of every converged run — 2,634 equations from 326 runs across 16 topologies — and resolved intermediate variables so that each formula can be checked as mathematics. Each equation was then checked automatically at two levels. First, device consistency: every transistor an equation references, whether through an explicit lookup or through parameter naming, must exist in that run's netlist; all 2,634 equations pass. Second, structural form: whether the equation has the correct functional dependence for that specification on that topology, with the device mapping grounded in the specific netlist — for example, GBW computed from the input-pair transconductance over the compensation capacitor on two-stage cells, and gain as the product of the stage gains. 84.9% of the corpus (2,236 equations) matches these canonical forms exactly. The remaining 15.1% are concentrated in a handful of algebraically valid idioms that the deliberately conservative matcher does not credit — for example, gain written as the product of named stage-gain variables, or formulas wrapped in type casts. We reviewed these by class; none triggers the audit's error patterns (wrong compensation capacitor, non-input-pair transconductance, target echo). Outside those idiom classes, one equation in the corpus is genuinely deficient: a GBW prediction that algebraically reduces to its own target. Notably, the sizing in that run had applied the standard design equation correctly — only the reported prediction echoed the design choice back, making it circular and uninformative rather than wrong about the circuit; had it masked a real shortfall, the next round's measurement would have exposed and corrected it. As a single occurrence in 2,634 equations, we regard it as an anomaly the audit itself surfaced.

Finally, as a control for pretrained-formula recall, two benchmark netlists (2SMC-N and FC-N at 45 nm) were obfuscated: device indices permuted, net names replaced by neutral tokens, and all topology identifiers removed, leaving only connectivity. On the disguised 2SMC-N the model identified all 8 device roles correctly; on the disguised FC-N, all 18. Both converged in the baseline number of rounds (Supplementary Table S2), and their equations passed the same two-level audit as the named topologies: of the 24 disguised-cell equations, 22 match canonical forms outright, and the remaining two fall into the same idiom classes seen corpus-wide.

## VI. Discussion

### *A. Why the Framework Works*

The framework succeeds by decomposing analog sizing into three complementary tasks, each handled by the component best suited to it. The LLM provides structural circuit understanding: fresh, topology-specific equations derived from each netlist give it a causal model of how specifications relate to device dimensions — the anchor that keeps sizing adjustments convergent rather than oscillatory, distinguishes the framework from LLM-as-optimizer approaches, and enables generalization across topologies without per-circuit modification. The calibration loop provides parametric accuracy that the LLM cannot achieve from first principles, and the prediction-error feedback provides robustness against the residual numerical error of first-order analysis — effects the closed-form expressions only partially capture — that persists after calibration. Together, these three components explain why the pace of convergence is set by the width of each cell's design window (Sec. V-C) rather than by prediction accuracy (Sec. V-B): convergence rests on correct structure and measured feedback, not on numerical equation quality.

### *B. Architectural Robustness and Structural Correctness*

While the deterministic calibration layer bounds the impact of parametric errors — measured values override analytical estimates at every round — the framework does require structural correctness: an equation predicting the wrong direction of influence would drive the design away from the target rather than toward it. The requirement itself distinguishes the approach from black-box methods: the LLM must understand which devices govern each specification, not merely search a parameter space. Sec. V-F verifies this property across the converged corpus, and the cross-model results show the remaining dependence directly: under the identical framework, reliability falls with model tier (Sec. V-E, Table V) — o3, for instance, closes the two James-comparison cells reliably yet no trials on the two hardest circuits.

Running each round as an independent call with no conversation history is a deliberate methodological choice — the most demanding condition for the model, in which convergence must rest entirely on the information provided by

TABLE VII LLM-DERIVED EQUATIONS FOR THE 2SMC-N MEDIAN TRIAL (45 NM), FINAL ROUND

| **Spec** | **LLM-derived equation (final round)** | **Governing devices** | **Direction correct?** |
|---|---|---|---|
| $A_v$ | $A_v = g_{m1}(r_{o1} \parallel r_{o3}) \cdot g_{m8}(r_{o7} \parallel r_{o8})$ | M1/M2, M8 ($g_m$); M3/M4, M7 ($r_o$) | ✓ Increasing $g_{m1}$ or $g_{m8}$, or the $r_o$ of either stage, increases gain |
| GBW | $\mathrm{GBW} = f_u = g_{m1} / (2\pi \cdot C_c)$, $g_{m1} = (g_m/I_d)_{meas} \cdot I_{d1}$ | M1; $C_c$ | ✓ Increasing $g_{m1}$ or decreasing $C_c$ increases GBW |
| PM | $\mathrm{PM} = 90° - \tan^{-1}(f_u/f_{p2}) - [\tan^{-1}(f_u/f_{p3}) - \tan^{-1}(f_u/2f_{p3})] - \tan^{-1}(f_u/f_{pRz}) - \tan^{-1}(f_u/f_{zf}) + \tan^{-1}(f_u/f_z) + \Delta\mathrm{PM}_{cal}$<br>$f_{p2} = g_{m8}C_c / 2\pi[C_1C_2 + C_c(C_1+C_2)]$;<br>$f_{p3} = g_{m3}/2\pi C_{22}$ (mirror pole, zero at $2f_{p3}$); $f_{pRz} = 1/2\pi R_zC_1$;<br>$f_z = f_u/(C_c/C_1)^{1/2}$; $f_{zf} = g_{m1}/2\pi C_{gd,in}$ (feedforward) | M8 ($f_{p2}$); M3/M4 ($f_{p3}$); M1 ($f_{zf}$); $R_z$ ($f_z$) | ✓ Raising $f_{p2}$ (via $g_{m8}$) or $f_{p3}$ (via $g_{m3}$) reduces phase loss; $R_z$ places the zero $f_z$ to recover phase |
| SR+ | $\mathrm{SR+} = K \cdot I_{tail}/C_c$, $K = 1.30$ (measured) | M5 (tail); $C_c$ | ✓ Increasing $I_{tail}$ or decreasing $C_c$ increases positive slewing |
| SR− | $\mathrm{SR-} = \min[f_{tail} \cdot I_{tail}/C_c, 0.9 \cdot I_7/(C_2 + C_{gd,in} + C_c)]$, $f_{tail} = 0.215 + 5.0 \cdot (V_{ds5} - V_{dsat5})$ | M5 (tail); M7 (sink); $C_c$; $C_L$ | ✓ More tail headroom raises $f_{tail}$; more $I_7$ raises the sink path; the lesser governs |
| Power | $P = (I_{ref} + I_{tail} + I_7) \cdot (V_{DD} - V_{SS})$; $I_{tail} = 0.92 \cdot m_5 \cdot I_{ref}$, $I_7 = 0.989 \cdot m_7 \cdot I_{ref}$ | M5, M7 mirrors | ✓ Scales with the mirror multiplicities $m_5$, $m_7$ and $I_{ref}$ that set every branch current |
| CMRR | $\mathrm{CMRR} = \mathrm{CMRR}_{meas} + 20 \cdot \log_{10}[(g_{m1} \cdot r_{o5})/(g_{m1} \cdot r_{o5})_{meas}]$ | M1; M5 (tail $r_o$) | ✓ Increasing $g_{m1} \cdot r_{o5}$ increases CMRR |
| Noise | $v_n = v_{n,meas} \cdot [½ \cdot g_{m1,meas}/g_{m1} + ½ \cdot (W_1L_1)_{meas}/(W_1L_1)]^{1/2}$ | M1 | ✓ Increasing $g_{m1}$ lowers the thermal term; increasing input-pair area $W_1L_1$ lowers the flicker term |
| Swing | $\mathrm{Swing} = (V_{DD} - V_{SS}) - (V_{dsat5} + 60\ \mathrm{mV}) - (V_{ov8} + 45\ \mathrm{mV})$ | M7; M8 | ✓ Lower $V_{dsat5}$ and $V_{ov8}$ widen the output window |
| ICMR | $\mathrm{ICMR}_{hi} = V_{DD} - (V_{th3} + V_{ov3}) + V_{gs1} - 0.10\ \mathrm{V}$; $\mathrm{ICMR}_{lo} = V_{SS} + V_{dsat5} + V_{gs1}$ | M1; M3 (diode); M5 (tail) | ✓ Larger $V_{ov3}$ lowers the upper limit; larger $V_{dsat5}$ raises the lower — smaller overdrives widen the range |
| THD | −19.0 dB | — | Measured value carried forward† |
| Offset | 0.5 mV | — | Measured value carried forward† |

Equations are written out from the `compute_sizing` of the design reported in Table III.
The subscript *meas* marks values read from the preceding rounds' simulations; the numeric constants are likewise measured anchors (the mirror ratios, the slewing factors $K$ and $f_{tail}$, and $\Delta\mathrm{PM}_{cal} = +2.8°$), and $C_1$, $C_2$ (incl. $C_L$), $C_{22}$, and $C_{gd,in}$ are node totals of measured device capacitances.
†Prediction carried forward from the previous round's measurement.

`calibrate()`, not on the LLM's reasoning continuity or incremental tuning. This demonstrates that convergence is a property of the framework architecture; a conversational mode with cumulative history could reduce round counts but is not required.

### C. Scalability, Runtime, and Practical Deployment

The framework sizes circuits from 6 transistors (single-stage) through 16 (the three-stage NMC, the most complex compensation architecture tested) to 30 (the complementary class-AB opamp, with even its bias transistors sized through LLM-derived equations), all with the same prompt structure. It operates unchanged across all six process nodes, the one-shot calibration capturing node-specific parameters automatically; per-cell convergence appears in Tables III and IV.

The deterministic components — `build_prompt()`, sizing execution, `calibrate()`, and the surrounding orchestration code — are fully reproducible and execute in seconds, as does each round's NGspice (or Cadence) bench-suite simulation. The non-deterministic component is the LLM's equation generation, roughly 300 lines per round; for the strongest model on the benchmark the budget is LLM-dominated, with a median response time of about 12 minutes per round and a median run wall time of 27 minutes. Mean rounds to convergence range from 1.2 on the easiest cell to 7.4 on the hardest, with corresponding mean wall times from roughly ten minutes to just over three hours. The framework runs with any LLM accessible through a standard chat interface (API or web), and no persistent model state is required between rounds. Model capability sets the ceiling: in the cross-model study every model interprets the netlists and produces executable sizing functions, while convergence on the tightest design windows — where many coupled specifications must be satisfied jointly — separates the strongest models from the rest (Sec. V-E, Table V; per-model token and timing medians in Supplementary Table S5).

## VII. CONCLUSION

We have demonstrated a design automation framework for analog circuit sizing that combines LLM-derived analytical equations with one-shot DC OP calibration and prediction-error feedback. The framework achieves all target specifications on circuits ranging from 6 to 30 transistors across six process nodes, spanning eight topology families that include single-stage, two-stage, and three-stage compensation architectures — despite large initial prediction errors, showing that convergence rests on structurally correct equations and measured feedback rather than on numerical equation accuracy. In the matched-specification comparisons, convergence requires only a few simulations across all trials (Table IV). The calibrated equations provide an interpretable, auditable sizing rationale — the primary differentiator of this work — a property absent from both black-box optimization and existing LLM-based sizing methods. Future work includes extension to broader circuit classes. The demonstrated ability of a frontier LLM to interpret raw circuit netlists, analyze device roles and bias conditions, and derive structurally correct design equations — without training, fine-tuning, or per-topology prompt customization — suggests that LLMs can serve as circuit-aware design automation tools, complementing rather than replacing the optimization-based and knowledge-based methods that dominate the current landscape.

## ACKNOWLEDGMENT

AI tools were used to assist in manuscript preparation. The authors take full responsibility for the content of this paper.

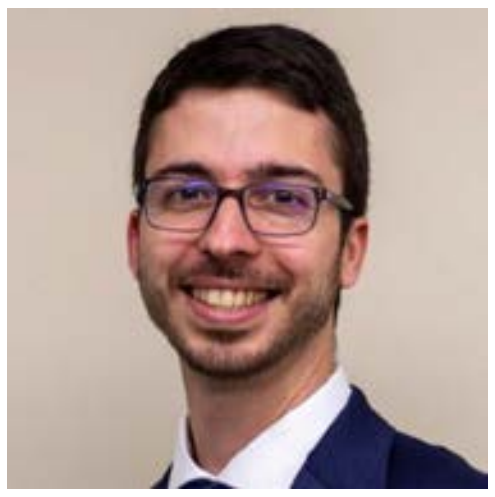

**Antonio J. Bujana** (Graduate Student Member, IEEE) received the B.S. degrees in engineering (electrical and biomedical) and the M.S. degree in engineering (electrical) from LeTourneau University, Longview, TX, USA, in 2015 and 2016, respectively. He is currently pursuing the Ph.D. degree in electrical engineering at Texas A&M University, College Station, TX, USA.

He has over 10 years of industry experience with Collins Aerospace. He is currently an Assistant Professor of electrical engineering at Benedictine College, Atchison, KS, USA. His research interests include large language model (LLM)-assisted circuit design automation.

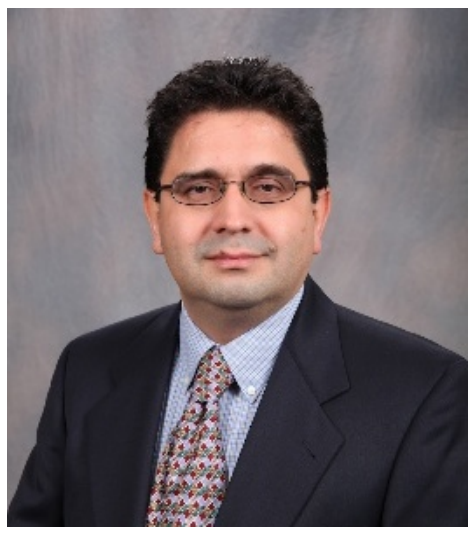

**Aydin I. Karsilayan** (Member, IEEE) received the B.S. and M.S. degrees in electrical engineering from Bilkent University, Ankara, Turkey, in 1993 and 1995, respectively, and the Ph.D. degree from Portland State University, Portland, OR, USA, in 2000.

He joined Texas A&M University, College Station, TX, USA, in 2000, where he is currently an Associate Professor in electrical and computer engineering. His current research interests include high-frequency analog filters, data converters, automatic tuning, mixed-mode integrated circuit design, RF communication circuits, and power harvesting. He has served as an Organizing Committee Member for the IEEE MWSCAS 2014. He has served as an Associate Editor for the IEEE Transactions on Circuits and Systems—I: Regular Papers, from 2002 to 2004, and the IEEE Transactions on Circuits and Systems—II: Express Briefs, from 2006 to 2010.

# A Self-Calibrating Framework for Analog Circuit Sizing Using LLM-Derived Analytical Equations: Supplementary Material

Antonio J. Bujana[1,2], *Graduate Student Member, IEEE*, and Aydin I. Karsilayan[1], *Member, IEEE*

This document contains Supplementary Tables S1–S6 referenced in the main manuscript.

TABLE S1 MATCHED-SPECIFICATION COMPARISON WITH PRIOR METHODS (FULL PER-SPEC DETAIL)

| Design | Av (dB) | GBW/UGF (MHz) | PM (deg) | Power (mW) | CMRR (dB) | Noise (nV/rtHz) | Offset (mV) | THD (dB) | Swing (Vpp) | ICMR (V) | Area (µm²) |
|---|---|---|---|---|---|---|---|---|---|---|---|
| Target [2] 2SMC-N | 45 | 20 | 45 | 0.2 | - | - | - | - | - | - | - |
| This work 45 PTM | 78.8±4.0 | 36.6±5.5 | 78.1±3.8 | 0.06±0.01 | - | - | - | - | - | - | - |
| This work 40 TSMC* | 47.9±1.6 | 32.2±8.5 | 87.5±4.5 | 0.06±0.01 | - | - | - | - | - | - | - |
| James [2] 45 TSMC | 48.3 | 21 | 46.1 | 0.15 | - | - | - | - | - | - | - |
| Target [2] FC-N | 45 | 20 | 45 | 0.2 | - | - | - | - | - | - | - |
| This work 45 PTM | 66.8±3.3 | 52.1±12.8 | 88.7±1.6 | 0.07±0.01 | - | - | - | - | - | - | - |
| This work 40 TSMC* | 46.3±1.2 | 70.1±26.3 | 85.4±2.1 | 0.09±0.02 | - | - | - | - | - | - | - |
| Target [8] FC-N | 60 | 1 | - | - | 80 | - | - | - | - | - | - |
| This work 90 PTM | 71.3±3.7 | 16.7±5.9 | - | - | 96.9±5.1 | - | - | - | - | - | - |
| LLM-USO [8] | 63 | 2.18 | - | - | 85.2 | - | - | - | - | - | - |
| Target [12] NMC | 85 | 0.7 | 55 | 0.05 | - | 500 | 3 | - | - | - | 250 |
| This work 180 PTM | 106±4.6 | 1.0±0.2 | 82.3±15.7 | 0.04±0.00 | - | 60.9±4.8 | 0.02±0.03 | - | - | - | 87.7±15.3 |
| This work 180 TSMC* | 117±1.3 | 0.9±0.1 | 70.1±10.4 | 0.05±0.00 | - | 203.9±49 | 0.15±0.13 | - | - | - | 118±42.5 |
| Atelier [12] 180 SMIC | 102 | 7.57 | 66.5 | 0.037 | - | 146 | 0.095 | - | - | - | 49.8 |
| Target [6] @180 | 65 | 10 | 50 | 10 | 100 | - | 1 | −26 | 1.71 | 1.71 | - |
| This work | 83.9±3.6 | 12.5±2.4 | 71.9±10.4 | 0.98±0.32 | 106.5±2.0 | - | 0.00±0.00 | −51.5±3.0 | 1.76±0.01 | 1.76±0.01 | - |
| EEsizer [6] | 64.0/62.5 | 12.6/50.1 | 53.6/49.5 | 9.73/5.10 | 127/120 | - | 0.16/0.46 | −26.0/−27.3 | 1.72/1.71 | 1.75/1.72 | - |
| Target [6] @130 | 65 | 10 | 50 | 10 | 100 | - | 1 | −26 | 1.23 | 1.23 | - |
| This work | 86.7±7.1 | 15.6±2.5 | 83.7±1.4 | 2.28±0.82 | 108.7±5.9 | - | 0.01±0.00 | −52.8±2.1 | 1.27±0.01 | 1.27±0.01 | - |
| EEsizer [6] | 62.5 | 10 | 47.7 | 3.9 | 139 | - | 0.03 | −24.8 | 1.72 | 1.75 | - |
| Target [6] @90 | 65 | 10 | 50 | 10 | 100 | - | 1 | −26 | 1.14 | 1.14 | - |
| This work | 87.2±6.7 | 24.1±4.2 | 76.0±2.5 | 1.33±0.71 | 106.3±3.9 | - | 0.00±0.00 | −56.1±1.9 | 1.17±0.00 | 1.17±0.00 | - |
| EEsizer [6] | - | - | - | - | - | - | - | - | - | - | - |

*TSMC PDK, Cadence Spectre, web-interface flow (Sec. IV-D); convergence counts in Table IV, per-trial outcomes in Table S3.

Aggregate entries are mean ± std over converged trials (N = 5; N = 3 for the TSMC rows).

Dashes mark specifications outside that comparison's matched target set or values the source does not report.

The James [2] row lists the specification values [2] reports for its two-stage design; [2] reports no specification values for its folded-cascode design, so that block carries no [2] row.

Atelier [12] reports its best result across 10 runs.

EEsizer [6] uses the same PTM node family, and supplies match except at 130 nm, where [6] retains 1.8 V while this work uses the node's 1.3 V supply; the swing/ICMR targets follow the same 0.95 × supply rule at every node.

EEsizer [6] succeeds in 2/5 trials at 180 nm (listed as x / y), 1/5 at 130 nm (its one success listed), and 0/5 at 90 nm, whose row therefore carries no values; this work converges 5/5 at all three nodes (rounds in Table IV).

TABLE S2 PERTURBATION, CAPACITANCE, SESSION, AND OBFUSCATION PROBES

| **Probe** | **Cell** | **Model** | **Conv** | **Rounds μ±σ** | **GBW \|err\|% r=0 (med)** | **GBW \|err\|% r≥1 (med)** |
|---|---|---|---|---|---|---|
| PERT00 (baseline) | 2SMC-N 45nm | gpt-5.6-sol | 5/5 | 2.4±0.5 | 473.3 | 20.4 |
| PERT05 | 2SMC-N 45nm | gpt-5.6-sol | 3/3 | 2.0±0.0 | 414.3 | 68.4 |
| PERT10 | 2SMC-N 45nm | gpt-5.6-sol | 3/3 | 2.3±0.6 | 506.6 | 27.9 |
| PERT20 | 2SMC-N 45nm | gpt-5.6-sol | 3/3 | 2.7±1.2 | 549.7 | 21.2 |
| DCAP (caps OFF) | 2SMC-N 45nm | gpt-5.6-sol | 5/5 | 2.4±0.9 | 536.1 | 10.5 |
| PERT00 (baseline) | FC-N 45nm | gpt-5.6-sol | 5/5 | 3.8±2.7 | 524.1 | 41.0 |
| PERT05 | FC-N 45nm | gpt-5.6-sol | 3/3 | 3.0±1.0 | 429.9 | 68.4 |
| PERT10 | FC-N 45nm | gpt-5.6-sol | 3/3 | 4.3±2.3 | 384.0 | 130.1 |
| PERT20 | FC-N 45nm | gpt-5.6-sol | 3/3 | 2.7±0.6 | 514.7 | 30.4 |
| DCAP (caps OFF) | FC-N 45nm | gpt-5.6-sol | 5/5 | 2.8±1.3 | 339.6 | 10.3 |
| Session: API fresh (baseline) | GBFC-P 65nm | claude-opus-5 | 5/5 | 7.4±3.0 | 23.2 | 26.8 |
| Session: web UI, cumulative chat | GBFC-P 65nm | claude-opus-5-web-cumulative | 4/5 | 4.0±0.8 | 22.8 | 14.4 |
| OBF baseline | 2SMC-N 45nm | claude-opus-5 | 5/5 | 2.0±0.0 | 498.4 | 11.9 |
| OBF XA7 (true 2SMC) | XA7 | claude-opus-5 | 3/3 | 2.0±0.0 | 471.1 | 14.9 |
| OBF XB3 (true FC) | XB3 | claude-opus-5 | 3/3 | 2.0±0.0 | 469.2 | 4.8 |

Probes: PERTxx = the four calibration factors corrupted independently (random sign, fixed ±xx% magnitude, per device and parameter) before each round.

DCAP = the prompt's per-device capacitance table (the $C_{gs}$, $C_{gd}$, $C_{db}$, $C_{sb}$ values of Sec. III-B) withheld.

Session = one independent API call per round (baseline) versus a single cumulative web chat.

OBF = the obfuscated netlists of Sec. V-F (XA7 = disguised 2SMC-N, XB3 = disguised FC-N).

The PERT00 rows are the matched full-framework baselines for the PERT and DCAP arms.

The two GBW columns give the median absolute relative prediction error at round 0 (uncalibrated initial estimates) and over the post-calibration rounds ($r \geq 1$), pooling every round of every run in the row.

The corrupted factors and the withheld capacitances enter only from round 1 — round 0 uses no calibration data — so the PERT and DCAP rows share their baselines' round-0 task, and similar $r = 0$ medians confirm nothing else differs.

Medians are reported because per-row samples are small and occasionally contain a single far-off intermediate design (individual rounds can exceed 100%).

Rounds entries are mean ± std over converged trials.

TABLE S3 CADENCE WEB-INTERFACE REPRODUCTIONS, PER TRIAL

| Run folder | Model | Outcome | Rounds |
|---|---|---|---|
| OTA_CAD_A1_DS_FC_P_180nm | claude-opus-4-6-web | CONVERGED | 7 |
| OTA_CAD_A1_DS_FC_P_180nm_1 | claude-opus-4-6-web | CONVERGED | 17 |
| OTA_CAD_A1_DS_FC_P_180nm_2 | claude-opus-4-6-web | NOT_CONVERGED | 20 |
| OTA_CAD_A1_DS_FC_P_180nm_3 | claude-opus-4-6-web | NOT_CONVERGED | 20 |
| OTA_CAD_A1_DS_FC_P_180nm_4 | claude-opus-4-6-web | CONVERGED | 16 |
| OTA_CAD_A1_DS_FC_P_180nm_5 | claude-opus-5-web | CONVERGED | 3 |
| OTA_CAD_A1_DS_FC_P_180nm_6 | claude-opus-5-web | CONVERGED | 1 |
| OTA_CAD_A1_DS_FC_P_180nm_7 | claude-opus-5-web | CONVERGED | 3 |
| OTA_CAD_A1_DS_FC_P_180nm_8 | claude-opus-5-web | CONVERGED | 4 |
| OTA_CAD_A1_DS_FC_P_180nm_9 | claude-opus-5-web | CONVERGED | 2 |
| OTA_CAD_Atalier_DS_NMC_N_180nm | claude-opus-5-web | CONVERGED | 2 |
| OTA_CAD_Atalier_DS_NMC_N_180nm_1 | claude-opus-5-web | CONVERGED | 2 |
| OTA_CAD_Atalier_DS_NMC_N_180nm_2 | claude-opus-5-web | CONVERGED | 2 |
| OTA_CAD_JAMES_DS_2SMC_N_40nm | claude-opus-5-web | CONVERGED | 2 |
| OTA_CAD_JAMES_DS_2SMC_N_40nm_1 | claude-opus-5-web | CONVERGED | 1 |
| OTA_CAD_JAMES_DS_2SMC_N_40nm_2 | claude-opus-5-web | CONVERGED | 2 |
| OTA_CAD_JAMES_DS_FC_N_40nm | claude-opus-5-web | CONVERGED | 3 |
| OTA_CAD_JAMES_DS_FC_N_40nm_1 | claude-opus-5-web | CONVERGED | 2 |
| OTA_CAD_JAMES_DS_FC_N_40nm_2 | claude-opus-5-web | CONVERGED | 2 |
| OTA_WEB_A1_DS_GBFC_P_65nm | claude-opus-5-web-cumulative | CONVERGED | 3 |
| OTA_WEB_A1_DS_GBFC_P_65nm_1 | claude-opus-5-web-cumulative | NOT_CONVERGED | 20 |
| OTA_WEB_A1_DS_GBFC_P_65nm_2 | claude-opus-5-web-cumulative | CONVERGED | 5 |
| OTA_WEB_A1_DS_GBFC_P_65nm_3 | claude-opus-5-web-cumulative | CONVERGED | 4 |
| OTA_WEB_A1_DS_GBFC_P_65nm_4 | claude-opus-5-web-cumulative | CONVERGED | 4 |

NOT_CONVERGED rows list the exhausted round budget (20).

The OTA_CAD_JAMES and OTA_CAD_Atalier blocks (the James [2] and Atelier [12] comparisons, at 40 nm and 180 nm) are the TSMC rows of Table IV.

The OTA_CAD_A1 FC-P 180 nm block is a standalone extended study — the FC-P benchmark cell rerun on the TSMC PDK through the web interface; it appears in neither Table III (its FC-P row is the PTM benchmark) nor Table IV (the anchor comparisons).

The OTA_WEB block is the cumulative-chat arm of Table S2's session probe; its fresh-per-round baseline is the API benchmark cell (GBFC-P 65 nm) and does not appear here.

The converged runs of this table are exactly the runs pooled by the web-flow rows of Table S5 (14 + 3 + 4 = 21).

TABLE S4 EQUATION-VERIFICATION SUMMARY (PER TOPOLOGY AND CORPUS TOTALS)

| Topology | Equations checked | Correct | Correct % | Incorrect | Not auto-matched |
|---|---|---|---|---|---|
| 1S-N | 60 | 45 | 75.0 | 0 | 15 |
| 1S-P | 540 | 486 | 90.0 | 0 | 54 |
| 2SMC-N | 324 | 272 | 84.0 | 1 | 51 |
| 2SMC-P | 60 | 53 | 88.3 | 0 | 7 |
| CM-CASC-N | 60 | 51 | 85.0 | 0 | 9 |
| CM-CASC-P | 60 | 53 | 88.3 | 0 | 7 |
| CM-N | 132 | 108 | 81.8 | 0 | 24 |
| CM-P | 60 | 51 | 85.0 | 0 | 9 |
| 30T | 415 | 354 | 85.3 | 0 | 61 |
| FC-N | 315 | 263 | 83.5 | 0 | 52 |
| FC-P | 156 | 128 | 82.1 | 0 | 28 |
| GBFC-N | 60 | 47 | 78.3 | 0 | 13 |
| GBFC-P | 312 | 253 | 81.1 | 0 | 59 |
| NMC-N | 56 | 50 | 89.3 | 0 | 6 |
| XA7 | 12 | 11 | 91.7 | 0 | 1 |
| XB3 | 12 | 11 | 91.7 | 0 | 1 |
| CORPUS TOTALS | 2634 | 2236 | 84.9 | 1 | 397 |
| DEVICE CONSISTENCY (pass) | 2634 | 2634 | 100.0 | 0 | — |
| DISGUISED CELLS (canonical) | 24 | 22 | 91.7 | 0 | 2 |

Correct = the equation matches the canonical structural form for that specification on that topology, with the device mapping grounded in the run's netlist (Sec. V-F).

Incorrect = fails the structural check outright; the corpus's single instance is the self-referential GBW prediction discussed in Sec. V-F.

Not auto-matched = equations in algebraically valid idioms that the deliberately conservative matcher does not credit; reviewed by class, none triggers the audit's error patterns.

In each topology row, every checked equation falls into exactly one category, so Correct + Incorrect + Not auto-matched = Equations checked.

Device consistency: every transistor an equation references exists in that run's netlist — all 2,634 equations pass.

The DISGUISED CELLS row subtotals the XA7 and XB3 rows (the obfuscated cells of Sec. V-F), which are already included in the corpus totals.

30T appears as FC2SRRIO in the audit records.

TABLE S5 PER-MODEL TOKEN AND TIMING SUMMARY (MEDIANS PER CONVERGED RUN)

| Model | Runs | In tok/rd | Visible out tok/rd | Reasoning tok/rd | LLM min/rd | Wall min/run |
|---|---|---|---|---|---|---|
| Claude Opus 5 | 166 | 7,412 | 7,638 | 36,721 | 9.0 | 59 |
| GPT-5.6 | 97 | 5,013 | 6,397 | 13,295 | 5.2 | 85 |
| Claude Opus 4.8 | 20 | 7,858 | 5,013 | 25,102 | 6.6 | 15 |
| o3 | 14 | 8,670 | 2,242 | 18,664 | 2.7 | 32 |
| Claude Opus 5 (web) | 14 | - | - | - | 11.1 | - |
| Claude Haiku 4.5 | 8 | 8,923 | 6,107 | 8,766 | 2.5 | 32 |
| Claude Opus 5 (web, cumul.) | 4 | - | - | - | 14.5 | - |
| Claude Opus 4.6 (web) | 3 | - | - | - | 8.7 | - |

Each model's row pools every converged run of that model — benchmark, anchor, and probe arms alike; Runs is that count.

Dashes mark the manual web-interface flows: the web interface reports no token metadata, and their wall times include operator relay, so neither is listed.

Sec. VI-C quotes benchmark-only medians for the primary model, Claude Opus 5 — about 12 min per round and 27 min per run — computed on the 60 benchmark runs alone (12 cells × 5 trials). The Claude Opus 5 row in this table instead pools all 166 of that model's converged runs across the benchmark, anchor, and probe arms, which is why its medians differ.

TABLE S6 PER-EQUATION AUDIT OF THE TABLE VII EXEMPLAR RUN (2SMC-N MEDIAN TRIAL)

| Spec | LLM-derived equation (final round) | Traces to devices | Verdict |
|---|---|---|---|
| Av | `20.0 * math.log10(max(gm1 * R1 * gm8 * R2, 1.0))` | M1,M3,M7,M8 | Correct (canonical-ref) |
| GBW | `GM_ID1 * Id1 / (2.0 * math.pi * min(max(Cc, 5e-14), 3e-13))` | M1 | Correct (canonical-intent) |
| PM | `PM = 90.0 - ph_p2 - ph_mir - ph_rz - ph_zf + ph_z + PM_CAL; min(max(PM, 0.0), 89.0)` | — | Correct (canonical-intent) |
| SR+ | `1.3 * (Itail / Cc) * 1e-06` | — | Correct (canonical-intent) |
| SR− | `min(min(max(SRN_A + SRN_B * (Vds5 - Vdsat5), 0.15), 0.95) * (Itail / Cc), 0.9 * (I7 / max(C2_slew + Cc, 1e-18))) * 1e-06` | M5,M7 | Correct (canonical-intent) |
| Power | `(3.9e-06 + 0.92 * M_5 * IREF + 0.989 * M_7 * IREF) * (VDD - VSS) * 1000.0` | M5,M7 | Correct (canonical-intent) |
| CMRR | `CMRR_dB = 46.35 + 20.0 * math.log10(max(gm1 * ro5, 1e-6) / (240e-6 * 37.7e3)); min(max(CMRR_dB, 20.0), 65.0)` | M5 | Correct (canonical-ref) |
| Noise | `200.6 * math.sqrt(0.5 * (0.00024 / max(GM_ID1 * Id1, 1e-09)) + 0.5 * (4e-05 * 4.5e-08 / max(W1 * M_1 * 4.5e-08, 1e-18)))` | M1 | Correct (canonical-intent) |
| Swing | `VDD - VSS - (0.078 + 0.06) - (Vov8 + 0.045)` | M8 | Correct (canonical-intent) |
| ICMR | `icmr_hi = V_net22 + Vgs1 - 0.10; icmr_lo = VSS + Vdsat5 + Vgs1; V_net22 = VDD - (Vth3 + Vov3); max(icmr_hi - icmr_lo, 0.05)` | M3 | Correct (canonical-ref) |
| THD | `-19.0` | — | Correct (anchored-const) |
| Offset | `0.5` | — | Correct (anchored-const) |

This table is the audit record for the exemplar run OTA_A1_DS_2SMC_N_45nm_3 — the same design reported in Tables III and VII.

Each row shows one final-round equation exactly as the Sec. V-F audit pipeline stored it, in the run's own Python, with its bound guards (min/max clamps) retained; where the stored form is only a guard around an intermediate (PM, CMRR, ICMR), the run's defining assignment is shown with it. Remaining intermediates are left unexpanded here. Table VII presents the same equations with every intermediate expanded for reading. Verdicts are assigned on the resolved mathematics: the audit expands all intermediates during checking, even where the stored form shown here retains them (Sec. V-F).

Verdict classes: canonical-ref = canonical form with parameters read from the measured per-device data; canonical-intent = canonical form with parameter values the model chose or anchored itself; anchored-const = a measured value carried forward as the prediction (the † rows of Table VII).

'Traces to devices' lists the transistors the audit resolved from the stored form; a dash marks stored forms with no direct per-device parameter (their device mappings appear in Table VII).